\documentclass{article}%
\usepackage{graphicx}
\usepackage{amsmath}
\usepackage{amsfonts}
\usepackage{amssymb}
\usepackage{setspace}
\usepackage{subfigure}%
\providecommand{\U}[1]{\protect\rule{.1in}{.1in}}
\onehalfspace
\newtheorem{theorem}{Theorem}

\newenvironment{proof}[1][Proof]{\noindent\textbf{#1.} }{\ \rule{0.5em}{0.5em}}

\begin{document}

\title{Higher order assortativity in complex networks}
\author{Alberto Arcagni, Rosanna Grassi, Silvana Stefani\thanks{Department of
Statistics and Quantitative Methods, University of Milano-Bicocca, Milano,
Italy}, Anna Torriero\thanks{Department of Mathematics, Finance and
Econometrics, Catholic University, Milano, Italy}}
\date{}
\maketitle

\begin{abstract}
Assortativity was first introduced by Newman and has been extensively studied
and applied to many real world networked systems since then. Assortativity is
a graph metrics and describes the tendency of high degree nodes to be directly
connected to high degree nodes and low degree nodes to low degree nodes. It
can be interpreted as a first order measure of the connection between nodes,
i.e. the first autocorrelation of the degree-degree vector. Even though
assortativity has been used so extensively, to the author's knowledge, no
attempt has been made to extend it theoretically. This is the scope of our
paper. We will introduce higher order assortativity by extending the Newman index
based on a suitable choice of the matrix driving the connections. Higher order
assortativity will be defined for paths, shortest paths, random walks of a given
time length, connecting any couple of nodes. The Newman assortativity is achieved for each of these measures when the matrix is the adjacency
matrix, or, in other words, the correlation is of order 1. Our higher order
assortativity indexes can be used for describing a variety of real networks,
help discriminating networks having the same Newman index and may reveal new
topological network features.

\textbf{Keywords:} Assortativity, Degree correlation, Networks, Paths, Random Walks

\textbf{PACS:} 89.75.Hc, 89.75.Fb, 02.50.Ga

\end{abstract}

\section*{Highlights}

\begin{itemize}
\item The concept of assortativity is extended to couples of nodes that are
not necessarily adjacent but connected through paths, walks, random walks.

\item A closed formula is found that can be tailored for the higher order
assortativity that is of interest for any application at hand.

\item The Newman assortativity is shown to be a particular case of all the
assortativity measures we propose.

\item Simulations show that our approach can help in revealing different
topologies in networks with the same Newman assortativity index.
\end{itemize}

\section{Introduction}

The concept of assortativity was first introduced by Newman (\cite{Newman
2002}) and has been studied and applied extensively to any kind of network
since. As is known, a network is said to be assortative if nodes of a
certain degree tend to be connected to nodes of the same degree, for instance, high degree nodes tend to be connected to high degree nodes and low
degree nodes tend to be connected to low degree nodes. A network is said to be
disassortative if high degree nodes tend to be connected to low degree nodes.
Quoting Newman, ``Many networks show ``assortative mixing'' on their degrees,
i.e. a preference for high degree vertices to attach to other high degree
vertices. Others show disassortativity mixing - high degree vertices attach
to low degrees ones''. Again ``Models that do not take it into account will
necessarily fail to reproduce correctly many of the behaviors of real-world
networked systems''.

We do agree with Newman that assortativity is an important measure and for
that we propose an extension that we may call higher order assortativity.

Assortativity is a graph metrics, i.e. related to the topology of the network
and is obtained by the Pearson coefficient of the degree-degree correlation
vectors. In fact, by definition, the Newman assortativity coefficient focuses
on the degree correlation of adjacent nodes, that is, the first autocorrelation
of the degree distribution. Based on the adjacency matrix, it can be
interpreted as a first order measure of the connection between nodes. The
measure is easy to apply and may give useful hints on the network topology
(for an extensive review on assortativity see \cite{noldusVanMeigen 2015}). On
the other hand, examples are given in the literature, where networks with
apparent different topologies show the same assortativity index, or
viceversa, networks with the same apparent topology show different
assortativity index (see \cite{Estrada 2011}, \cite{Alderson 2005 a},
\cite{Alderson 2005 b}). This would imply somehow that assortativity is not a
\textquotedblleft good\textquotedblright\ measure, but of course it comes from
its relative simplicity and in our opinion it can be efficiently used for a
preliminary inspection to discover the network topology. A way of making
assortativity more \textquotedblleft efficient\textquotedblright\ is to extend
it theoretically and this is the core of our paper. In fact, our original
contribution is meant to extend assortativity to higher order autocorrelations
between nodes. We provide a unified approach to assortativity, extending it to
higher order connections based on paths, shortest paths and random walks, through suitable
definitions of the matrix governing the connections. We provide a closed
formula, suitable for all cases we consider. Our formula may be used for
measuring assortativity between two not necessarily adjacent nodes but
connected through paths, shortest paths or random walks. We show that the Newman
assortativity index comes out as a particular case for all the measures we
propose. Through simulations we will show that our approach helps in finding
synthetic indicators for revealing the network topology. Finally, we provide
some hints about choosing the most suitable higher order indicator according
to the network at hand and to which flow we are interested to discover.
Essentially, extending the concept of assortativity, that is, understanding
how high degree nodes are connected to high degree nodes directly or through
paths or walks, may help to better understand networks and their topology. As
a consequence, if we understand the patterns of interactions, we may leverage
this knowledge to improve the flow of knowledge and information (\cite{anklam
2003}). The paper is structured as follows: after the literature review on
extensions of assortativity and preliminaries on graph theory (Section 2), we
discuss how Newman assortativity is obtained as the first autocorrelation of
the degree-degree vector (Subsection 3.1). In the following, higher order
assortativity is introduced on random walks (Section 4), paths and geodesics
(Section 5). Simulations and conclusions follow.

\subsection{Literature review}

A short literature review on assortativity mixing has been studied extensively
since the paper by \cite{Newman 2002}. The original definition of
assortativity has been given in \cite{Newman 2002} for an unweighted and
undirected network. Examples of assortativity applied to many real complex
networks can also be found in \cite{Newman 2002}: physics coauthorship,
biology coauthorship, film actor collaborations, company directors are
examples of assortative networks, while Internet, WWW, protein interaction and
neural networks are examples of disassortative networks. Random graphs and
Barabasi-Albert networks are examples of non-assortative networks. For a
complete and recent review of assortativity we refer to \cite{noldusVanMeigen
2015}. More generally, in F. da Costa \textit{et al.} (\cite{Da Costa 2007}) a survey of
measures of complex networks is discussed. Assortativity is sometimes called
homophily, when referring specifically to social networks (\cite{Mayo 2015}).
Among the metrics related to Newman assortativity, we quote the Zagreb
Index or S(G) (\cite{Alderson 2005 a}; \cite{Grassi 2010}). Alderson \textit{et al.}
(\cite{Alderson 2005 a}) discuss the S(G) metric and scale-free graphs. They
provide examples of networks having the same degree sequence with an apparent
different topology and introduce S(G) as indicators for discriminating network
topologies. In \cite{orsiniEtAl 2015} many measures are used and accompany
assortativity. The authors employ the $dk-$series, a set of characteristics of
the network topology, to study the statistical dependencies between different
network properties. Through the $dk-$series, they study the average degree,
assortativity, clustering and so on. From an empirical point of view,
\cite{Mayo 2015} developed and executed an algorithm to evaluate degree
correlation between nodes separated by more than one step. Through shortest
paths, they studied three online social networks and compared their long range
degree correlation behavior to those of three non-social networks by measuring
both the average number of neighbors and calculating the Pearson correlation
score. Authors conclude that results are not clear cut and require further
investigation. Many other measures are used and accompany assortativity, such as
in \cite{orsiniEtAl 2015}.

A discussion on how appropriate the Pearson coefficient is for comparing mixing patterns in networks of a different size can be found in \cite{Litvak2}. They show that the Pearson coefficient in scale-free networks decreases with the network size, thus making impossible to compare, for example, two web crawls of different sizes. Alternatively, they suggest a degree-degree dependency measure based on Spearman's rho \cite{Litvak1}.

Although assortativity is such an important feature and so popular in the
complex networks world, as far as the authors know, no attempts have been made
to extend such a measure to a theoretical and unifying view.

\section{Basics about graph theory}

We quickly recall some standard definitions about graph theory. We will assume
familiarity with basic theoretical concepts (see \cite{Harary}, \cite{hogg 2005}, \cite{mood 1974}). A network is a graph $G=\left(  V,E\right)  $, where
$V=\left\{  v_{1},v_{2},...,v_{n}\right\}  $ is the set of vertices and
$E\subseteq V\times V$ the set of edges (or links). Let us denote with
$\left\vert V\right\vert =n$ and $\left\vert E\right\vert =m$ \ the
cardinality of the sets $V$ and $E$, respectively. We consider simple graphs,
i.e without loops and multiple edges. An undirected graph is a graph in which
if $(u,v)\in E,$ then $(v,u)\in E$. When two vertices share a link, they are
called adjacent. The degree $d_{i}$ of a vertex $v_{i}$ $(i=1,...,n)$ is the
number of edges incident with it. We denote by $\mathbf{d}^{T}=\left[
\begin{array}
[c]{cccc}%
d_{1} & d_{2} & \ldots & d_{n}%
\end{array}
\right]  $ the degree sequence of the graph and $\mathbf{D}=diag(d_{i})$. A
walk is a sequence of adjacent vertices $v_{1},v_{2},...,v_{l}.$ A $u-v$ path
is a walk connecting $u$ and $v$ in which all vertices are distinct. A
shortest path joining vertices $u$ and $v$ is called a $u-v$ geodesic. The
distance $dist\left(  u,v\right)  $ between two vertices $u$ and $v$ is the
length of the $u-v$ geodesic; the diameter $D$ of $G$ is the maximum of
$dist\left(  u,v\right)  ,$ $u,v\in V.$ A graph is connected if for each pair
of vertices $u$ and $v$ there is a path connecting $u$ and $v$. A graph is
$k-$ regular if every vertex has the same degree $k$. A graph is bipartite if
$V$ can be divided into two separate sets $V_{1}$ and $V_{2}$ such that every
node in $V_{1}$ and $V_{2}$ is not connected to each other.

A non-negative $n-$square matrix $\mathbf{A}=\left[  a_{ij}\right]  ,$
$\left(  i,j=1,2,...,n\right)  ,$ representing the adjacency relationships
between vertices of $G,$ is associated to the graph (the adjacency matrix);
the off-diagonal elements $a_{ij}$ of $\mathbf{A}$ are equal to $1$ if
vertices are adjacent, 0 otherwise. Througout the paper we will assume that
all graphs are connected. The matrix $\mathbf{A}$ is said to be primitive if
there exists a positive integer $k$ such that $\mathbf{A}^{k}>\mathbf{0}$.

\section{From correlation to assortativity}

Let $\mathbf{x}^{T}=\left[
\begin{array}
[c]{cccc}%
x_{1} & x_{2} & \ldots & x_{n}%
\end{array}
\right]  $ and $\mathbf{y}^{T}=\left[
\begin{array}
[c]{cccc}%
y_{1} & y_{2} & \ldots & y_{n^{\prime}}%
\end{array}
\right]  $ be two real vectors and $\mathbf{E}=[e_{ij}],$ $1\leq i\leq n,1\leq
j\leq n^{\prime}$ a non-negative \emph{matrix of weights\footnote{In general,
the weights can be relative frequencies or probabilities.} }between the
couples $(x_{i},y_{j})$, such that $\mathbf{1}^{T}\mathbf{E1}=\sum_{i=1}%
^{n}\sum_{j=1}^{n^{\prime}}e_{ij}=1$ (where $\mathbf{1}$ is the unit vector).
Note that $n$ and $n^{\prime}$ are in general different, but in our case we
suppose $n$ equal to $n^{\prime}$.

The sums by columns of the matrix $\mathbf{E}$, $\mathbf{q}_{\mathbf{x}%
}=\mathbf{E}^{T}\mathbf{1}$, are the \emph{marginal weights} of the vector
$\mathbf{x}$ and the sums by rows of the matrix $\mathbf{E}$, $\mathbf{q}%
_{\mathbf{y}}=\mathbf{E1,}$ are the \emph{marginal weights }of the vector
$\mathbf{y}$.

In order to discuss the Newman assortativity index in the next section, it is
useful to recall the definition of the Pearson's linear coefficient (see
\cite{mood 1974}) and to adapt the notation to our context.

Let
\begin{align*}
\mu_{\mathbf{x}}  &  =\sum_{i}x_{i}q_{i}^{\mathbf{x}};\;\mu_{\mathbf{y}}%
=\sum_{j}y_{j}q_{j}^{\mathbf{y}};\\
\sigma_{\mathbf{x}}  &  =\sqrt{\sum_{i}(x_{i}-\mu_{\mathbf{x}})^{2}%
q_{i}^{\mathbf{x}}};\;\sigma_{\mathbf{y}}=\sqrt{\sum_{j}(y_{j}-\mu
_{\mathbf{y}})^{2}q_{j}^{\mathbf{y}}}%
\end{align*}
be respectively the weighted mean value of $\mathbf{x}$, and $\mathbf{y}$, the
weighted standard deviation of $\mathbf{x}$ and $\mathbf{y}$.

The Pearson's linear correlation index is defined by:
\begin{equation}
r(\mathbf{x}, \mathbf{y})=\frac{\sum_{i}\sum_{j}(x_{i}-\mu_{\mathbf{x}}%
)(y_{j}-\mu_{\mathbf{y}})e_{ij}}{\sigma_{\mathbf{x}}\sigma_{\mathbf{y}}}
\label{eq:rho_def}%
\end{equation}
and all the following expressions are equivalent:\bigskip%
\begin{align}
r(\mathbf{x}, \mathbf{y})  &  =\frac{\sum_{i}\sum_{j}x_{i}\,y_{j}\,e_{ij}%
-\mu_{\mathbf{x}}\mu_{\mathbf{y}}}{\sigma_{\mathbf{x}}\sigma_{\mathbf{y}}%
}=\frac{1}{\sigma_{\mathbf{x}}\sigma_{\mathbf{y}}}\left[  \mathbf{x}%
^{T}\,\mathbf{E}\,\mathbf{y}-\mu_{\mathbf{x}}\mu_{\mathbf{y}}\right]
=\nonumber\\
&  =\frac{\sum_{i}\sum_{j}x_{i}\,y_{j}\,(e_{ij}-q_{i}^{\mathbf{x}}%
q_{j}^{\mathbf{y}})}{\sigma_{\mathbf{x}}\sigma_{\mathbf{y}}}=\frac{1}%
{\sigma_{\mathbf{x}}\sigma_{\mathbf{y}}}\left[  \mathbf{x}^{T}\,\left(
\mathbf{E}-\mathbf{q}_{\mathbf{x}}\,\mathbf{q}_{\mathbf{y}}^{T}\right)
\,\mathbf{y}\right]  \label{eq:rho_mat_gen}%
\end{align}
where $\mathbf{q}_{\mathbf{x}}^{T}=\left[
\begin{array}
[c]{cccc}%
q_{1}^{\mathbf{x}} & q_{2}^{\mathbf{x}} & \ldots & q_{n}^{\mathbf{x}}%
\end{array}
\right]  $ and $\mathbf{q}_{\mathbf{y}}^{T}=\left[
\begin{array}
[c]{cccc}%
q_{1}^{\mathbf{y}} & q_{2}^{\mathbf{y}} & \ldots & q_{n^{\prime}}^{\mathbf{y}}%
\end{array}
\right]  $.

By definition, the Pearson's index is the ratio between the covariance of
$\mathbf{x}$ and $\mathbf{y}$ and the maximum absolute value that the
covariance can assume, i.e. $\sigma_{\mathbf{x}}\sigma_{\mathbf{y}}$ (see
formula (\ref{eq:rho_def})). Note that $r(\mathbf{x},\mathbf{y})=0$ occurs in
case of absence of correlation. Absence of correlation also occurs in the case
in which all the components of one of the two vectors are all equal. However,
in this case the ratio takes the form $0/0,$ then $r(\mathbf{x},\mathbf{y})$
is undefined.

Observe that, in probability theory (see \cite{hogg 2005}) $\mathbf{x}$ and
$\mathbf{y}$ represent two random variables with $\mathbf{E}$ the matrix of
their joint probability distribution. The value $e_{ij}$ is the probability to
observe the couple of values $(x_{i},y_{j})$. The vector $\mathbf{q}%
_{\mathbf{x}}$ is the marginal probability distribution of the variable
$\mathbf{x}$. The value $q_{i}^{\mathbf{x}}$ is the probability to observe the
value $x_{i}$ without considering the value assumed by the other variable.
Analogously, $\mathbf{q}_{\mathbf{y}}$ is the marginal probability
distribution of the variable $\mathbf{y}$ and $q_{j}^{\mathbf{y}}$ is the
probability to observe the value $y_{j}$ without considering the value assumed
by the other variable. Finally, let $\mathbf{D}_{\mathbf{q}_{\mathbf{y}}%
}=diag(q_{j}^{\mathbf{y}}),$ $j=1,..,n^{\prime};$ observe that, when all the
elements of the vector $\mathbf{q}_{\mathbf{y}}$ are positive, this matrix is
invertible. $\mathbf{P}_{\mathbf{x}|\mathbf{y}}=\mathbf{E\,D}_{\mathbf{q}%
_{\mathbf{y}}}^{-1}=[p_{ij}^{\mathbf{x}|\mathbf{y}}]=[e_{ij}/q_{j}%
^{\mathbf{y}}]$, $1\leq i\leq n,1\leq j\leq n^{\prime}$ is a stochastic by
column matrix that represents the partial distribution of $\mathbf{x}$
conditioned to the values assumed by the other variable. The value
$p_{ij}^{\mathbf{x}|\mathbf{y}}$ represents the probability to observe the
value $x_{i}$ assuming that the value $y_{j}$ has been observed.

Note that, if the variable $\mathbf{x}$ is \emph{independent} from the
variable $\mathbf{y}$ then $p_{ij}^{\mathbf{x}|\mathbf{y}}=q_{i}^{\mathbf{x}}
$ for all $1\leq i\leq n,1\leq j\leq n^{\prime}$. Independence is a symmetric
relation and it implies $\mathbf{E}=\mathbf{q}_{\mathbf{x}}\mathbf{q}%
_{\mathbf{y}}^{T}$ having rank equal one. From equation (\ref{eq:rho_mat_gen})
it follows that, if variables $\mathbf{x}$ and $\mathbf{y}$ are independent,
$\mathbf{E}-\mathbf{q}_{\mathbf{x}}\,\mathbf{q}_{\mathbf{y}}^{T}$ is a matrix
with entries all equal to zero, then $r(\mathbf{x}, \mathbf{y})=0. $

For later use, let us consider the particular case in which $\mathbf{x}%
=\mathbf{y}=\mathbf{d}$ and $\mathbf{E}$ a $n\times n$-symmetric matrix; then
we get:
\begin{align*}
\mu_{\mathbf{x}}=\mu_{\mathbf{y}}  &  :=\mu\\
\sigma_{\mathbf{x}}=\sigma_{\mathbf{y}}  &  :=\sigma\\
\mathbf{q}_{\mathbf{x}}=\mathbf{q}_{\mathbf{y}}  &  :=\mathbf{q}.
\end{align*}
and Pearson's linear correlation index is:
\begin{align}
r(\mathbf{d},\mathbf{d})  &  =\frac{\sum_{i}\sum_{j}d_{i}d_{j}\,e_{ij}-\mu
^{2}}{\sigma^{2}}=\frac{1}{\sigma^{2}}\left[  \mathbf{d}^{T}\,\mathbf{E}%
\,\mathbf{d}-\mu^{2}\right]  =\nonumber\\
&  =\frac{\sum_{i}\sum_{j}d_{i}\,d_{j}\,(e_{ij}-q_{i}q_{j})}{\sum_{j}d_{j}%
^{2}q_{j}-\sum_{i}\sum_{j}d_{i}\,d_{j}\,q_{i}q_{j}}=\frac{\mathbf{d}%
^{T}\,\left(  \mathbf{E}-\mathbf{q}\,\mathbf{q}^{T}\right)  \,\mathbf{d}%
}{\mathbf{d}^{T}\,(\mathbf{D}_{\mathbf{q}}-\mathbf{q}\,\mathbf{q}%
^{T})\,\mathbf{d}} \label{Pearson coeff}%
\end{align}
where $\mathbf{D}_{\mathbf{q}}$ is the diagonal matrix, with diagonal entries
equal to the elements of vector $\mathbf{q}$.

\subsection{Newman's assortativity index}

A network is assortative if high degree nodes tend to be connected to
high degree nodes, whereas it is disassortative if high degree nodes tend to
be connected to low degree nodes. The definition of assortativity was first
introduced by Newman (\cite{Newman 2002}) using the Pearson's coefficient of
the degree-degree correlation in an unweighted and undirected network.

Let $G=(V,E)$ be un undirected and unweighted graph, with degree sequence
$\mathbf{d}^{T}=\left[
\begin{array}
[c]{cccc}%
d_{1} & d_{2} & \ldots & d_{n}%
\end{array}
\right]  $. Assuming in (\ref{Pearson coeff}) $\mathbf{E}=\frac{1}%
{2m}\mathbf{A},$ we get $\mathbf{q}=\frac{1}{2m}\mathbf{d,}$ $\mathbf{D}%
_{\mathbf{q}}=\frac{1}{2m}\mathbf{D}$ and the formula (\ref{Pearson coeff})
gives the well known measure proposed by Newman, that can be rewritten as:%

\begin{equation}
\rho=\frac{\mathbf{d}^{T}\left(  \frac{\mathbf{A}}{2m}-\frac{\mathbf{d}%
\,\mathbf{d}^{T}}{4m^{2}}\right)  \,\mathbf{d}}{\mathbf{d}^{T}\,(\frac
{\mathbf{D}}{2m}-\frac{\mathbf{d}\,\mathbf{d}^{T}}{4m^{2}})\,\mathbf{d}},
\label{eq:rho_A}%
\end{equation}

or equivalently as (see the Appendix):\bigskip%

\begin{equation}
\rho=\frac{\frac{1}{2m}\sum_{i}\sum_{j}d_{i}d_{j}a_{ij}-\left[  \frac{1}%
{4m}\sum_{i}\sum_{j}\left(  d_{i}+d_{j}\right)  a_{ij}\right]  ^{2}}{\frac
{1}{4m}\sum_{i}\sum_{j}\left(  d_{i}^{2}+d_{j}^{2}\right)  a_{ij}-\left[
\frac{1}{4m}\sum_{i}\sum_{j}\left(  d_{i}+d_{j}\right)  a_{ij}\right]  ^{2}}.
\label{Newman}%
\end{equation}
\bigskip

Note that (\ref{Newman}) is undefined when the graph $G$ is regular, since
numerator and denominator are both equal to zero (see \cite{Estrada 2011 b},
pag. 32), or in other words there is no variability within the degree sequence.

The Newman's assortativity coefficient focuses on the degree correlation
between only adjacent nodes, so it can be interpreted as a first order measure
of the connection between nodes. In this paper, the assortativity definition
will be extended also to nodes connected by random walks, paths and shortest
paths. All these alternative definitions can be modelled using our unified
approach through suitable definitions of matrix $\mathbf{E}.$ To this aim, the
formula (\ref{Pearson coeff}) can be used to measure the assortativity between
two not necessarily adjacent nodes but connected through random walks, paths
and shortest paths.

More in general, $\mathbf{E}$ can be a weight matrix of order $n$, expressing
any reciprocal relation between each couple of nodes, giving rise to different
indices of assortativity. In the following Sections, we will introduce higher
order assortativity for nodes connected through random walks, paths and
shortest paths. For each case, applications taken from the literature will be provided.

\section{Assortativity through random walks\label{sec:RW}}

In this Section a new measure of assortativity based on random walks of length
$l$ is introduced. We will show that Newman's assortativity index is only
a special case of our measure.

Given the graph $G=(V,E)$, let $E_{w,l}\subseteq V\times V$ be the set of the
undirected walks of length $l$. For $l=1$, $E_{w,1}=E$ is the set of the
edges. Let $\mathbf{E}_{w,l}$ be the matrix of the probabilities that a walk
randomly chosen from $E_{w,l}$ connects vertices $i$ and $j$. Putting
$\mathbf{E}=\mathbf{E}_{w,l},$ the formula (\ref{Pearson coeff}) can be written
as:
\begin{equation}
\rho_{w,l}=\frac{\mathbf{d}^{T}\left(  \mathbf{E}_{w,l}-\mathbf{q}%
\mathbf{q}^{T}\right)  \mathbf{d}}{\mathbf{d}^{T}\left(  \mathbf{D}%
_{\mathbf{q}}-\mathbf{q}\mathbf{q}^{T}\right)  \mathbf{d}}.
\label{Pearson coeff RW}%
\end{equation}
where $\rho_{w,l}$ denotes the linear Pearson coefficient of the degree
sequence with weights given by the matrix $\mathbf{E}_{w,l}$. Observe that for
$l=1$ $\mathbf{E}_{w,1}=\frac{1}{2m}\mathbf{A}$, $\mathbf{q}=\frac{1}%
{2m}\mathbf{d}$ so that $\rho_{w,1}$ matches with the Newman measure.

The following result concerns the asymptotic behavior of
(\ref{Pearson coeff RW}) (the proof is reported in the Appendix):

\begin{theorem}
Let $G=(V,E)$ be a graph with adjacency matrix $\mathbf{A}$ and degree
sequence $\mathbf{d.}$ Let $\mathbf{P}$ be the transition matrix of a Markov
chain on $G=(V,E)$. If $\mathbf{P}$ is primitive, the assortativity of order
$l,$ $\rho_{w,l}$ vanishes as \thinspace$l\rightarrow\infty.$
\end{theorem}

Observe that, for bipartite connected graphs, $\mathbf{P}$ is not primitive.
In this case, $\mathbf{P}^{l}$ does not converge to the stationary
distribution (see \cite{Noble Daniel 1977}, \cite{Lovasz}). Moreover,
similarly to the Newman assortativity (\ref{Newman}), if $G=(V,E)$ is a regular
graph, the assortativity of order $l$ ($\rho_{w,l}$) is undefined.

A possible application of higher order assortativity through random walks can
be found in an input-ouput network (see \cite{Blolch 2011}), where the
movement of goods between the sectors of an economy is modeled as a random
walk. Goods, like random walkers, start out at a given position and repeatedly
choose an edge incident to their current position. The choices are made
according to a probability distribution determined by the edge weights. The
goods proceed for an arbitrarily long time or until a prescribed goal is
reached. From formula (\ref{Pearson coeff RW}) we may be able to track down
the movements of those goods.\bigskip

\section{Assortativity through paths}

In this Section we define a new measure of assortativity based on paths of
length $l.$ The Newman assortativity measure will be extended by taking into
account not only the direct connection between two nodes (i.e. the adjacency)
but also the higher order neighborhood structure of the network through paths.

Being the graph $G$ connected, two nodes $i$ and $j$ are always linked by a
path of some length $l$. However, an assortativity measure should also capture
more complex structural features, such as the degree of the nodes belonging to
the $(i,j)$-path. This can be obtained by (\ref{Pearson coeff}) by a suitable
choice of the matrix $\mathbf{E}$ as explained in this Section.

\subsection{Assortativity through degree-based paths}

Given the graph $G=(V,E)$, let $E_{p,l}\subseteq V\times V$ be the set of
undirected paths of length $l$ between any couple of nodes. Let $\mathbf{E}%
_{p,l}$ be the weighted matrix associated to $E_{p,l}$, whose entries are
$e_{ij}=\frac{1}{e}\sum_{i_{1}i_{2}....i_{l-1}}\left(  d_{i_{1}}d_{i_{2}%
}....d_{i_{l-1}}\right)  $, $i,j=1,...,n,$ where $i_{1}i_{2}....i_{l-1}$ are
the nodes belonging to all $l$ - paths between $i$ and $j$ and $e$ is the sum
of $d_{i_{1}}d_{i_{2}}.....d_{i_{l-1}}$ over all the $l$ - paths in $E_{p,l}.$

Assuming $\mathbf{E}=\mathbf{E}_{p,l},$ formula (\ref{Pearson coeff})
becomes:
\begin{equation}
\rho_{p,l}=\frac{\mathbf{d}^{T}\left(  \mathbf{E}_{p,l}-\mathbf{q}%
\mathbf{q}^{T}\right)  \mathbf{d}}{\mathbf{d}^{T}\left(  \mathbf{D}%
_{\mathbf{q}}-\mathbf{q}\mathbf{q}^{T}\right)  \mathbf{d}}. \label{eq:rho_p}%
\end{equation}
\bigskip

A particular case occurs when $l=1,$ i.e. the path lengths are all equal to
$1$. Then $E_{p,1}$ becomes the set of edges $E,$ $\mathbf{E}_{p,1}=\frac
{1}{2m}\mathbf{A,}$ since $e$ reduces to the sum of the degrees ($2m).$ Vector
$\mathbf{q}$ becomes $\frac{1}{2m}\mathbf{d}$, so that $\rho_{p,1}$
corresponds to the Newman measure. Notice that, unlike assortativity through
random walks, (\ref{eq:rho_p}) exists only for $l$ lower or equal to the
length of the longest path.

Let $\mathbf{E}_{up,l}$ be the matrix associated to the set of undirected
paths of length $l$ obtained by $\mathbf{E}_{p,l}$ putting elements $d_{i_{k}%
}=1,$ where $i_{1}i_{2}....i_{l-1}$ are the nodes belonging to all $l$ - paths
between $i$ and $j$. The entries $e_{ij}$ of this matrix simply become the
number of the existing $l-$ path between nodes $i$ and $j,$ divided by $e$
(simply the number of all $l-$ paths in $E_{p,l}$).

Formula (\ref{eq:rho_p}) can be rewritten as:%
\begin{equation}
\rho_{up,l}=\frac{\mathbf{d}^{T}\left(  \mathbf{E}_{up,l}-\mathbf{q}%
\mathbf{q}^{T}\right)  \mathbf{d}}{\mathbf{d}^{T}\left(  \mathbf{D}%
_{\mathbf{q}}-\mathbf{q}\mathbf{q}^{T}\right)  \mathbf{d}}. \label{eq:rho_up}%
\end{equation}

In this case, assortativity extends the degree correlations beyond
adjacency through paths of a given length, but keeping into account only the
degree of the source and the target nodes.

This assortativity is related to a topological index, the \emph{higher-order
connectivity index}, extensively used in Chemistry (see \cite{Gutman 2013}).
The higher order connectivity index was proposed in the literature as a
generalization of the Randi\'{c} index (see \cite{Randic 1975}).

Following the definition found in \cite{Lu-Li 2004}, for an integer $l\geq1,$
the $l$ $-$ connectivity index is defined as:
\[
\chi_{\alpha}^{l}\left(  G\right)  =\sum_{\left(  i,j\right)  }\left(
d_{i}d_{i_{1}}d_{i_{2}}....d_{i_{l-1}}d_{j}\right)  ^{\alpha},
\]
$\alpha>0,$ where the sum runs over all $\left(  i-j\right)  $ paths of length
$l$ of $G$.

The higher-order connectivity index had various chemical applications, but so
far not many mathematical results have been obtained on $\chi_{\alpha}%
^{l}\left(  G\right)  $; some results can be found in \cite{Lu-Li 2004},
\cite{Rada Araujo 2002} and \cite{Yero Gutman 2010}.

Observe that in (\ref{eq:rho_p}) and (\ref{eq:rho_up}) the first term on the
numerator, i.e. $\mathbf{d}^{T}\mathbf{E}_{p,l}\mathbf{d}$ and $\mathbf{d}%
^{T}\mathbf{E}_{up,l}\mathbf{d}$, differs from the $l$-connectivity index when the exponent respectively is $\alpha=1$ and $\alpha=0$  by the multiplicative factor $\frac{1}{e}$.

\subsection{Assortativity through shortest paths}

From formula (\ref{Pearson coeff}), a measure of assortativity, based on
shortest paths of length $l \leq D$, can also be defined. Let $E_{sp,l}%
\subseteq V\times V$ the set of geodesics of length $l$ and $\mathbf{E}%
_{sp,l}$ be its associated matrix, whose entries $e_{ij}$ are defined as in
(\ref{eq:rho_up}) where $i_{1}i_{2}....i_{l-1}$ are the nodes belonging to all
$l$ - \textit{shortest paths} between $i$ and $j,$ divided by the cardinality
of $E_{sp,l}$. Formula (\ref{Pearson coeff}) becomes in this case:%
\[
\rho_{sp,l}=\frac{\mathbf{d}^{T}\left(  \mathbf{E}_{sp,l}-\mathbf{q}%
\mathbf{q}^{T}\right)  \mathbf{d}}{\mathbf{d}^{T}\left(  \mathbf{D}%
_{\mathbf{q}}-\mathbf{q}\mathbf{q}^{T}\right)  \mathbf{d}};
\]
observe that this index has been proposed by Mayo \textit{et al.} in \cite{Mayo 2015}.

Another assortativity measure, based on shortest paths, can be defined in the
following way.

Given the connected graph $G=(V,E)$, let $\mathbf{H}_{\alpha}=\left[
h_{ij}\right]  ,$ be the matrix having the diagonal entries equal to zero,
whereas $h_{ij}=dist(i,j)^{-\alpha}$ for $i\neq j$, $\alpha>0$ real parameter.
Assuming in (\ref{Pearson coeff}) $\mathbf{E}=\frac{1}{h}\mathbf{H}_{\alpha},$
where $h=\sum_{i}\sum_{j}h_{ij},$ we obtain:
\begin{equation}
\rho_{c,\alpha}=\frac{\mathbf{d}^{T}\left(  \frac{1}{h}\mathbf{H}_{\alpha
}-\mathbf{q}\mathbf{q}^{T}\right)  \mathbf{d}}{\mathbf{d}^{T}\left(
\mathbf{D}_{\mathbf{q}}-\mathbf{q}\mathbf{q}^{T}\right)  \mathbf{d}}%
=\frac{\frac{1}{h}\mathbf{d}^{T}\mathbf{H}_{\alpha}\mathbf{d}-\mathbf{d}%
^{T}\mathbf{q}\mathbf{q}^{T}\mathbf{d}}{\mathbf{d}^{T}\left(  \mathbf{D}%
_{\mathbf{q}}-\mathbf{q}\mathbf{q}^{T}\right)  \mathbf{d}}. \label{eq:rho_c}%
\end{equation}

Differently from the previous indices, $\rho_{c,\alpha}$ measures
assortativity also taking into account, in addition to the degree sequence,
the length of the shortest path between nodes $i$ and $j$, i.e. $dist\left(
i,j\right)  $. This measure generalizes Newman assortativity, including all couples of nodes (adjacent or not) in the formula but with decreasing weights as
the distance between them increases.

We can prove the following Theorem (see the Appendix 2):
\begin{theorem}
Let $G=(V,E)$ be a simple connected graph. The coefficient $\rho_{c,\alpha}$
tends to $\rho$ as $\alpha$ tends to infinity.
\end{theorem}

Observe that formula (\ref{eq:rho_c}) is related to another global network
indicator, known in the literature as clumpiness (see \cite{Estrada 2008}).
Given a connected graph $G=(V,E)$, the clumpiness coefficient is defined by:%

\[
\Lambda(G,\mathbf{d},\alpha)\mathbf{=}\overset{n(n-1)/2}{\underset{i>j}{\sum}%
}\frac{d_{i}d_{j}}{(dist\left(  i,j\right)  )^{\alpha}}=\frac{1}{2}%
\mathbf{d}^{T}\mathbf{H}_{\alpha}\mathbf{d.}%
\]

It is easy to observe that $\rho_{c,\alpha}$ can also be written as:
\[
\rho_{c,\alpha}=\frac{\frac{2}{h}\Lambda(G,\mathbf{d},\alpha)-\mathbf{d}%
^{T}\mathbf{q}\mathbf{q}^{T}\mathbf{d}}{\mathbf{d}^{T}\mathbf{D}_{\mathbf{q}%
}\mathbf{d}-\mathbf{d}^{T}\mathbf{q}\mathbf{q}^{T}\mathbf{d}}.
\]

Note that, as the authors point out in (\cite{Estrada 2008}), clumpiness
and the Newman's index measure different features of the network. Indeed, the
clumpiness index $\Lambda(G,\mathbf{d},\alpha)$ increases with the increase of
the node degrees but, on the contrary, it decreases with the increase in the
distance between them, and various examples of clumped assortative and clumped
disassortative networks are provided in (\cite{Estrada 2008}).

A classic example of a flow moving through geodesics in a network is given
by logistics. A driver delivering a package normally knows and selects the
shortest route possible, so that the package's trajectory follows geodesic
paths through the road network \cite{Borgatti 2005}. The higher order
assortativity based on shortest paths may help to track down the flow of packages.

\section{Simulations}

In this section some simulations are performed in order to analyse and compare
the different assortativity measures previously defined.

We simulate 100 graphs of same order $n=30$ with the same degree sequence:%
\[
\mathbf{d}=\left[  7^{(3)},6^{(2)},5^{(8)},4^{(3)},3^{(8)},2^{(5)}%
,1^{(1)}\right]  .
\]
Graphs are connected, non-isomorphic and without loops. They have same size
$m=60$, average degree $\mu=4$ and density $\delta=\frac{2m}{n(n-1)}=0.1379$,
but they differ by topology.

\begin{figure}[ptb]
\subfigure[Graph $G_{1}$]{\includegraphics[width=0.5\textwidth]{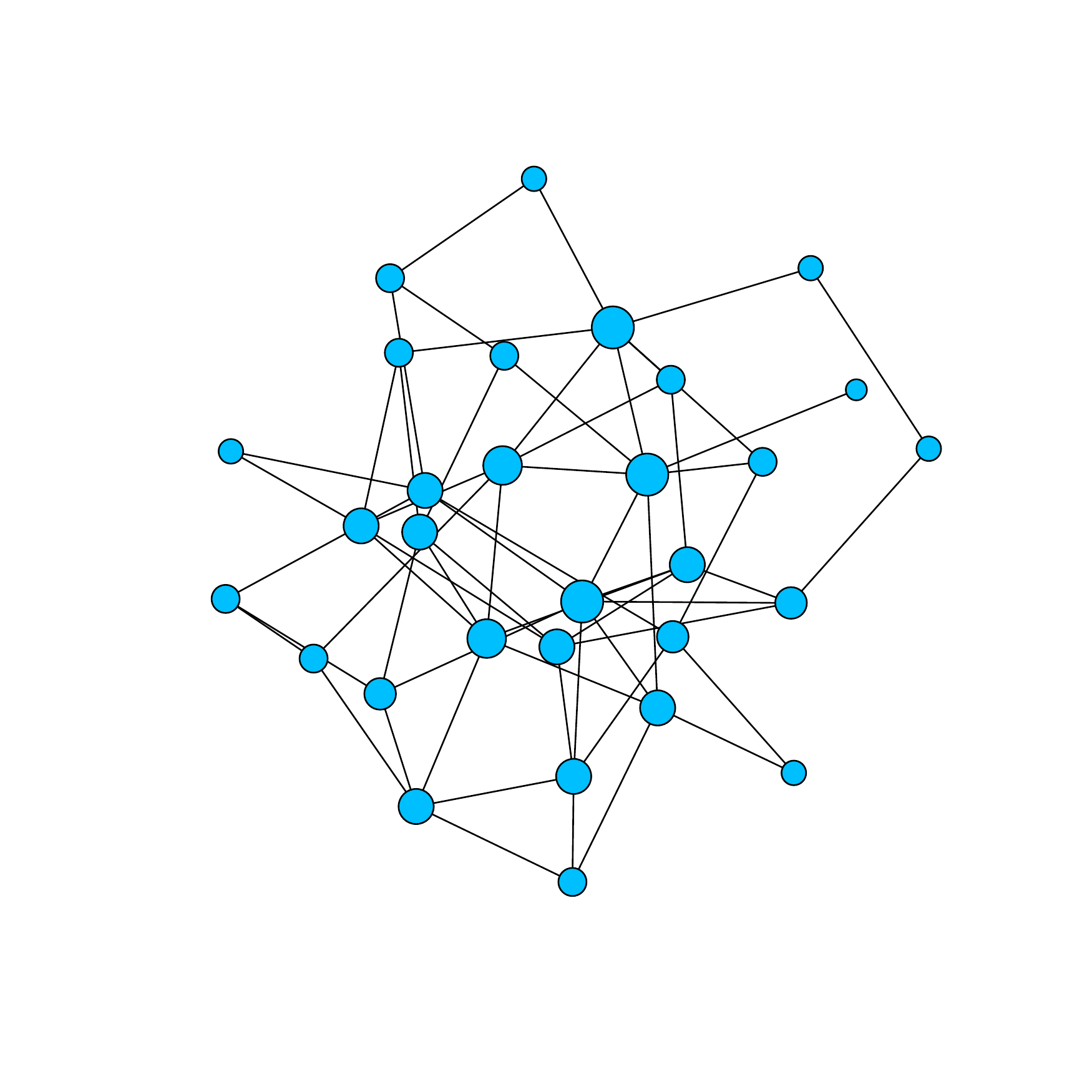}}\label{fig:G1}
\subfigure[Graph $G_{2}$]{\includegraphics[width=0.5\textwidth]{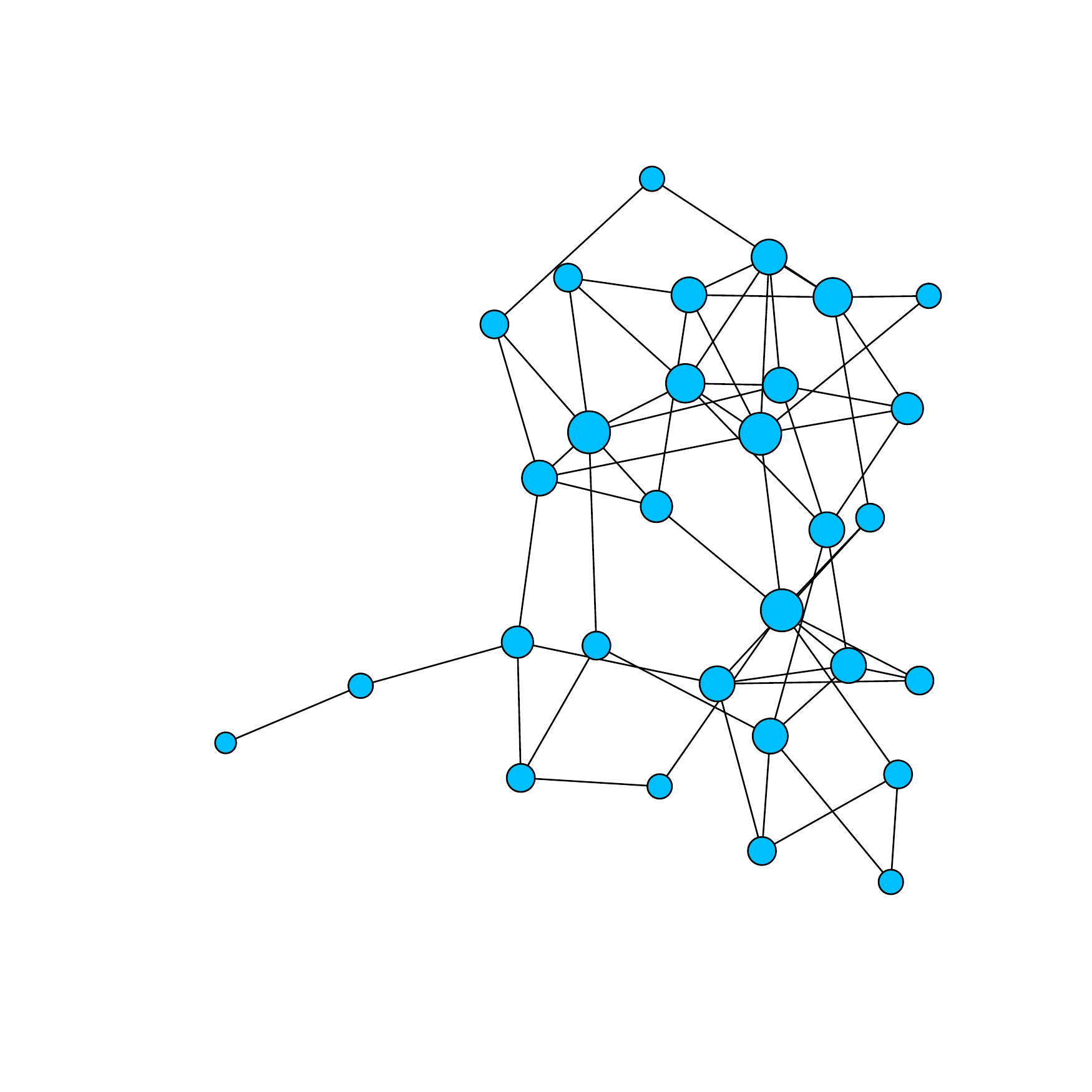}}\label{fig:G2}
\caption{Two simulated graphs with the same Newman's coefficient.}%
\label{fig:firstComparisonGraphs}%
\end{figure}

\begin{figure}[ptb]
\subfigure[Graph $G_{3}$]{\includegraphics[width=0.45\textwidth]{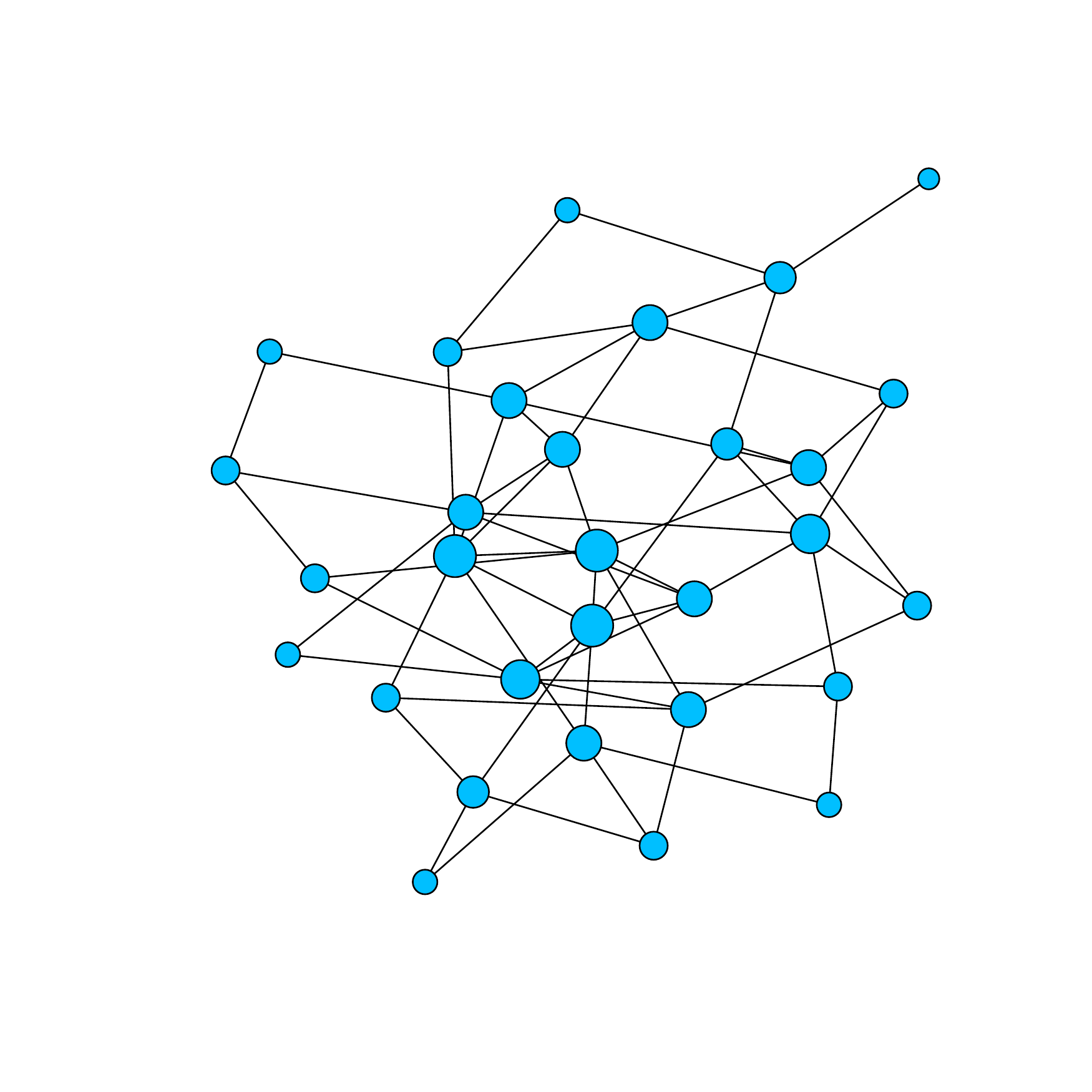}}\label{fig:G3}
\subfigure[Graph $G_{4}$]{\includegraphics[width=0.45\textwidth]{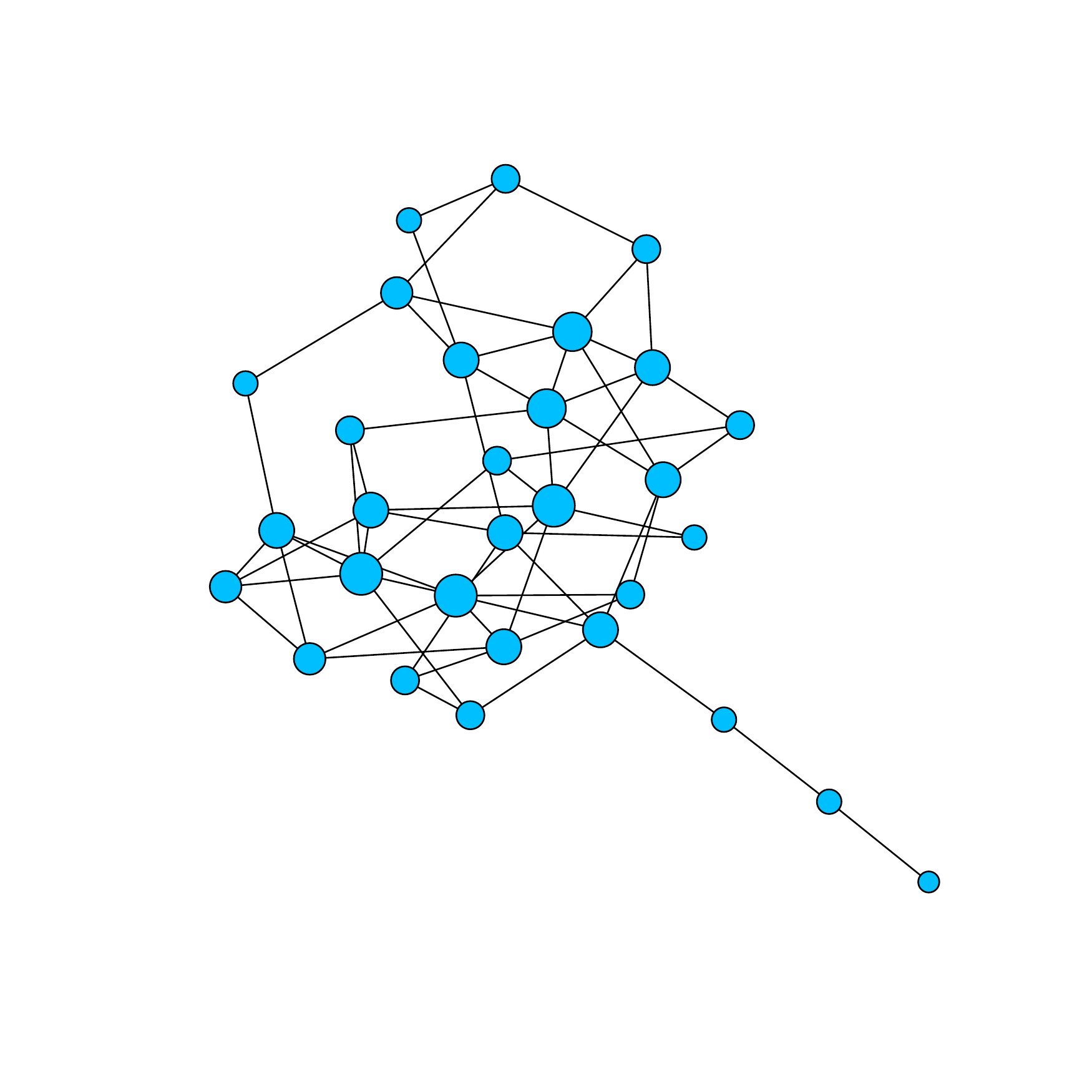}}\label{fig:G4}%
\caption{Two simulated graphs with the same Newman's coefficient.}%
\label{fig:secondComparisonGraphs}%
\end{figure}

Let us consider the simulated graphs, $G_{1},G_{2}$
and $G_{3},G_{4}$, respectively in Figures \ref{fig:firstComparisonGraphs} and
\ref{fig:secondComparisonGraphs} (higher degree nodes are thicker).

Graphs $G_{1}$ and $G_{2}$ have the same Newman's index, $\rho
(G_{1})=\rho(G_{2})=-0.0584$ and they are equally disassortative but, by
inspection of Figure \ref{fig:firstComparisonGraphs}, they are
topologically different. Indeed, we will see that other structural
parameters, based on shortest paths, give different values.

To this end we check some classical network indicators: the diameter $D,$ the
average (shortest) path length:%
\[
L=\frac{1}{n(n-1)}\sum_{i\not =j}dist(i,j)
\]
and the clustering coefficient, also named transitivity (see \cite{Newman
2001}):
\[
C=\frac{3\left\vert T(3)\right\vert }{P_{2}},
\]
where $T(3)$ is the number of triangles and $P_{2}$ is the number of 2-paths.

Observe that $D(G_{1})=4,$ whereas $D(G_{2})=6;$ $L(G_{1})=2.4483$ whereas
$L(G_{2})=2.6230$. $C(G_{1})=0.1091$ whereas $C(G_{2})=0.1773$.

Table \ref{tab:firstComparison} reports the assortativity measures
proposed in this paper\footnote{The number of undirected paths of length
$l=10$ is of order $10^{5}$ of magnitude and our R-code \cite{R code} uses a
recursive algorithm in order to extract the undirected paths, making the
procedure computationally intensive. For this reason, the measures depending
on lengths $l$ have been evaluated until $l=10$.}, referring to $G_{1}$ and
$G_{2},$ allowing us to analyze all the measures simultaneoulsy.

\begin{table}[ptb]
\caption{Higher order assortativity measures of graphs $G_{1}$ and $G_{2}$.}%
\label{tab:firstComparison}
\centering
\begin{tabular}
[c]{r|rrr|rrr}%
$l$ & $\rho_{w,l}(G_{1})$ & $\rho_{p,l}(G_{1})$ & $\rho_{up,l}(G_{1})$ &
$\rho_{w,l}(G_{2})$ & $\rho_{p,l}(G_{2})$ & $\rho_{up,l}(G_{2})$\\\hline
1 & -0.0584 & -0.0584 & -0.0584 & -0.0584 & -0.0584 & -0.0584\\
2 & 0.1826 & -0.0787 & -0.0864 & 0.3501 & 0.0951 & 0.1024\\
3 & -0.0322 & -0.0593 & -0.0563 & 0.0183 & -0.0181 & -0.0054\\
4 & 0.0719 & 0.0030 & 0.0010 & 0.1796 & -0.0661 & -0.0495\\
5 & -0.0245 & -0.0306 & -0.0316 & 0.0319 & -0.0348 & -0.0309\\
6 & 0.0347 & -0.0376 & -0.0336 & 0.1066 & -0.0494 & -0.0457\\
7 & -0.0169 & -0.0486 & -0.0403 & 0.0301 & -0.0549 & -0.0538\\
8 & 0.0185 & -0.0452 & -0.0390 & 0.0680 & -0.0532 & -0.0573\\
9 & -0.0110 & 0.0379 & -0.0334 & 0.0247 & -0.0497 & -0.0543\\
10 & 0.0105 & -0.0397 & -0.0361 & 0.0451 & -0.0489 & -0.0543
\end{tabular}
\end{table}

\begin{figure}[ptb]
\centering
\subfigure[Assortativity through random walks of $G_1$]{\includegraphics[width=0.4\textwidth]{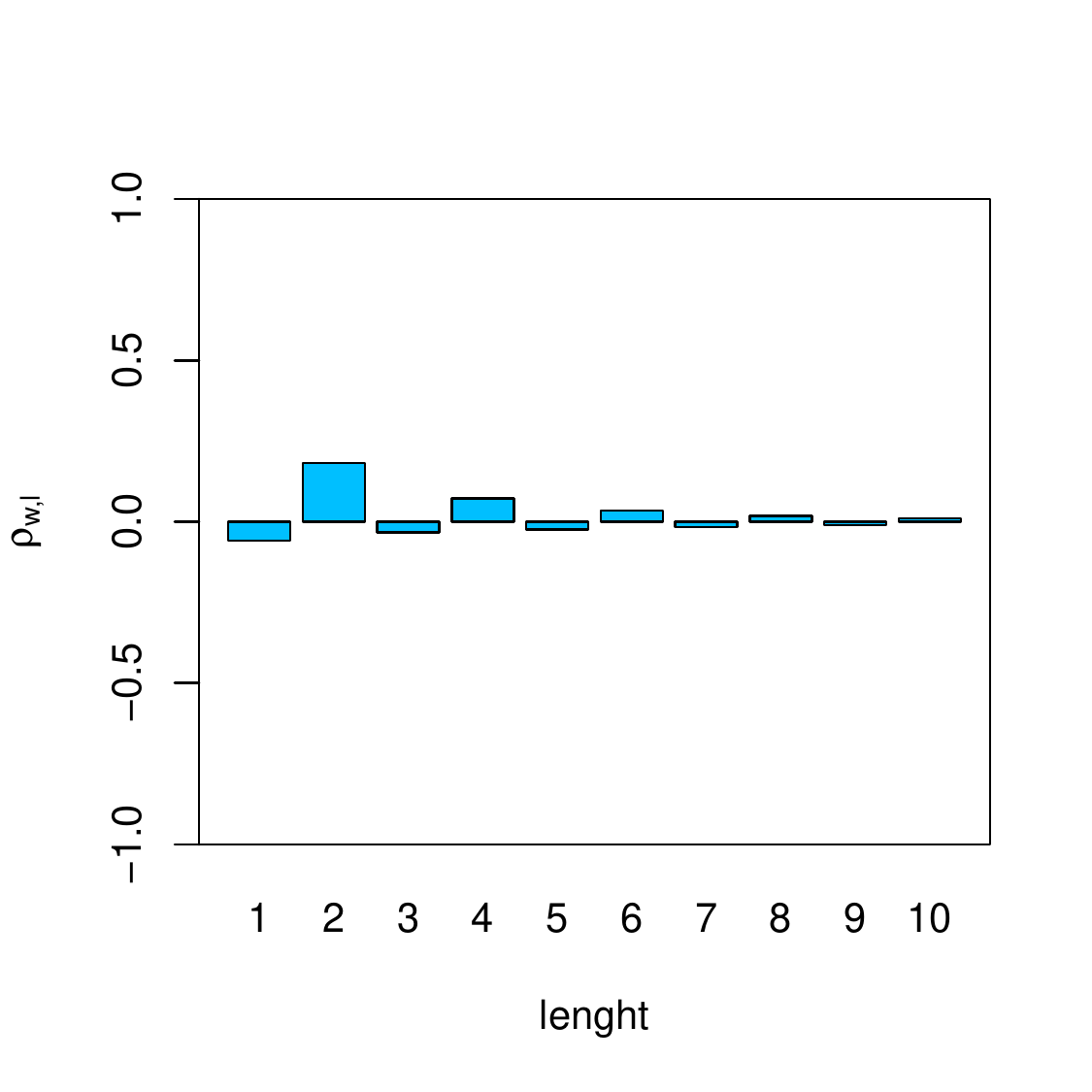}}\label{fig:G1RW}
\hspace{0.1\textwidth}
\subfigure[Assortativity through random walks of $G_2$]{\includegraphics[width=0.4\textwidth]{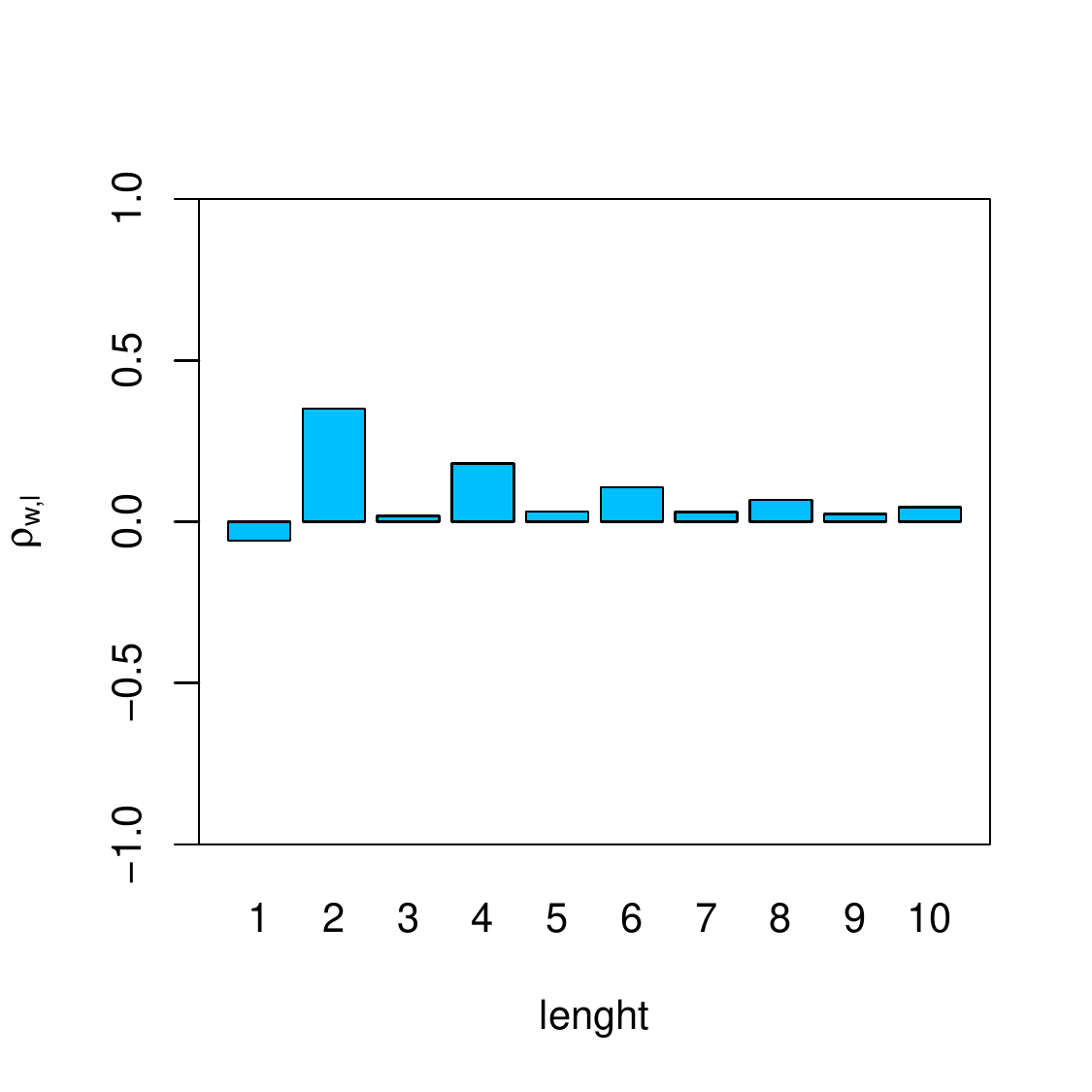}}\label{fig:G2RW}
\subfigure[Assortativity through degree-based paths of $G_1$]{\includegraphics[width=0.4\textwidth]{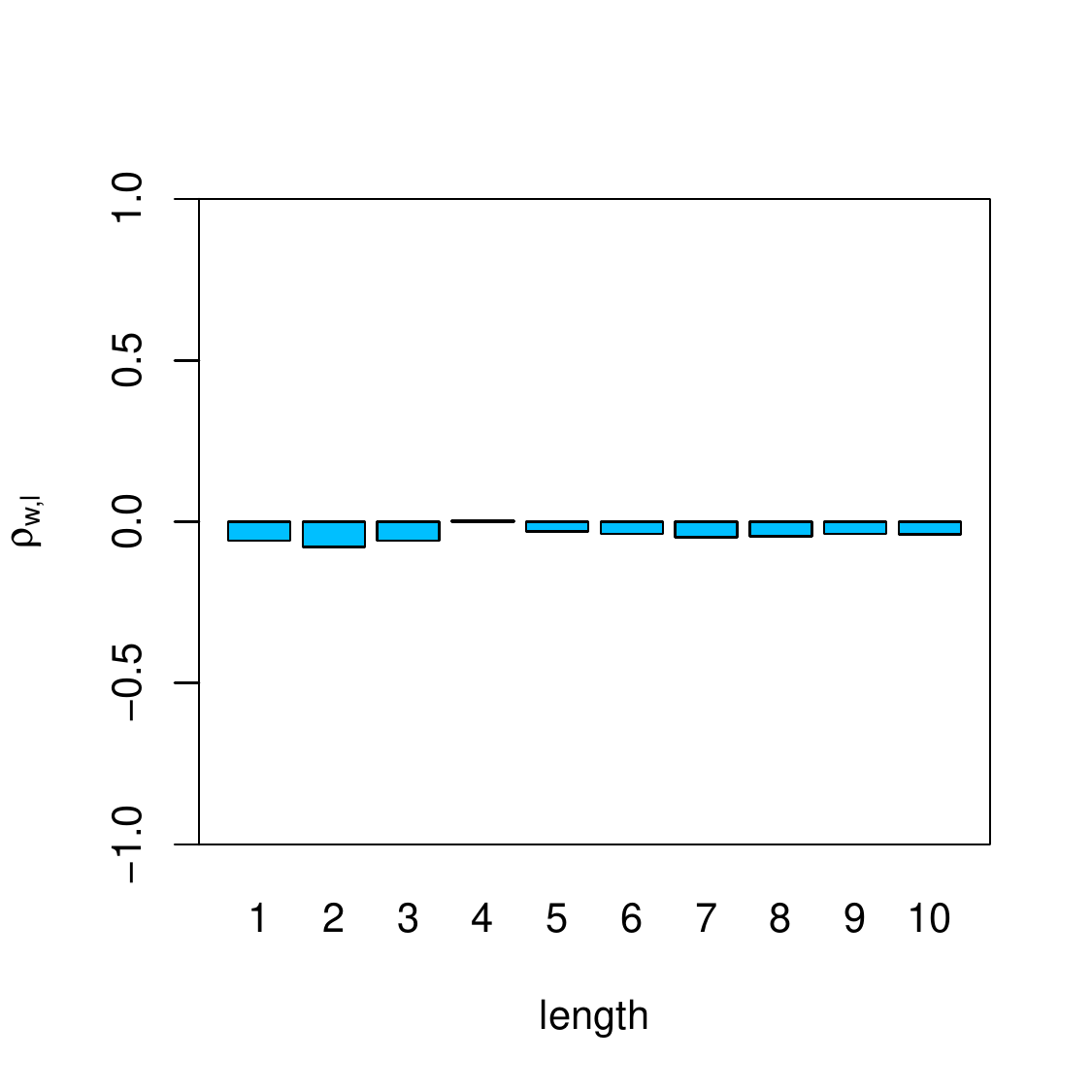}}\label{fig:G1WP}
\hspace{0.1\textwidth}
\subfigure[Assortativity through degree-based paths of $G_2$]{\includegraphics[width=0.4\textwidth]{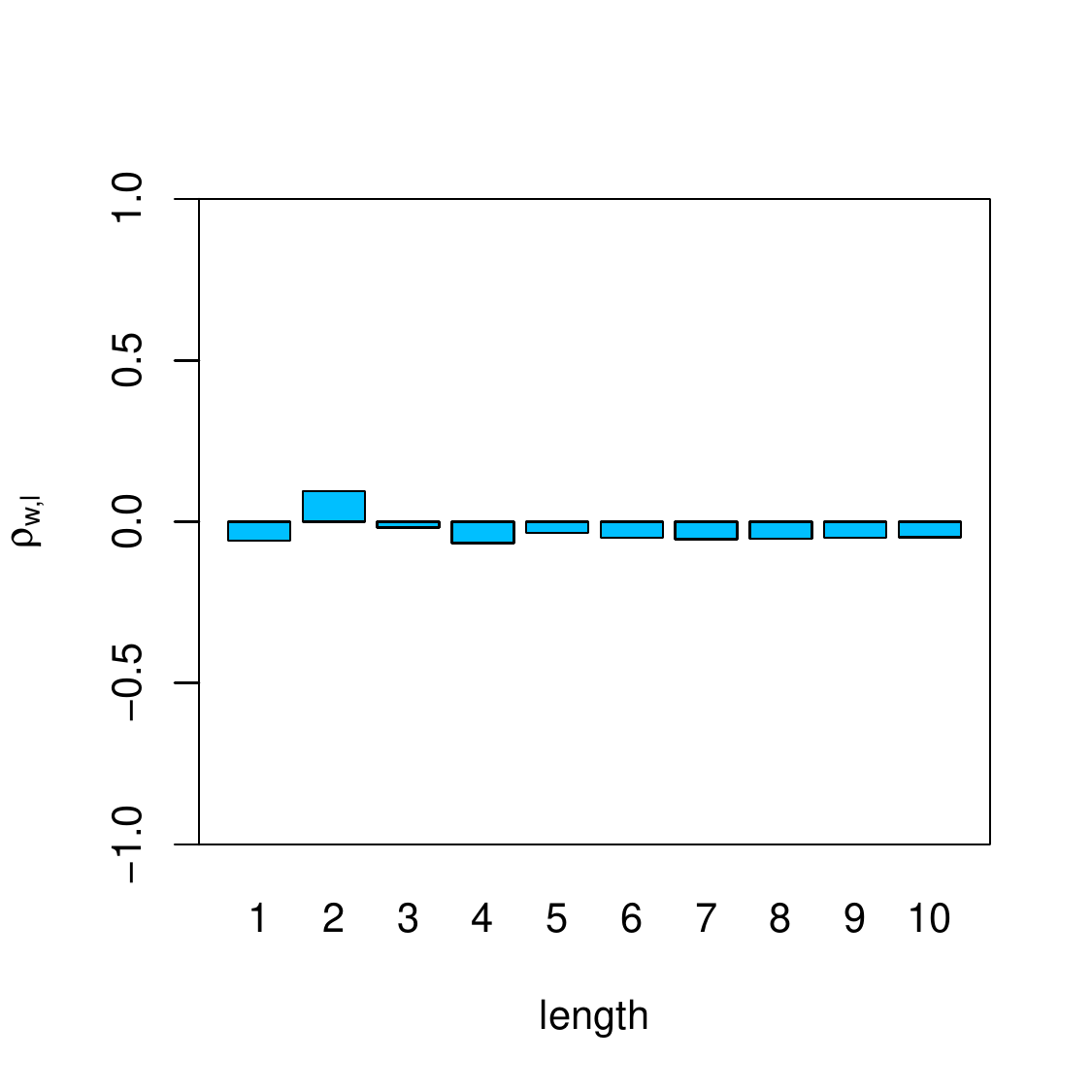}}\label{fig:G2WP}
\subfigure[Assortativity through paths of $G_1$]{\includegraphics[width=0.4\textwidth]{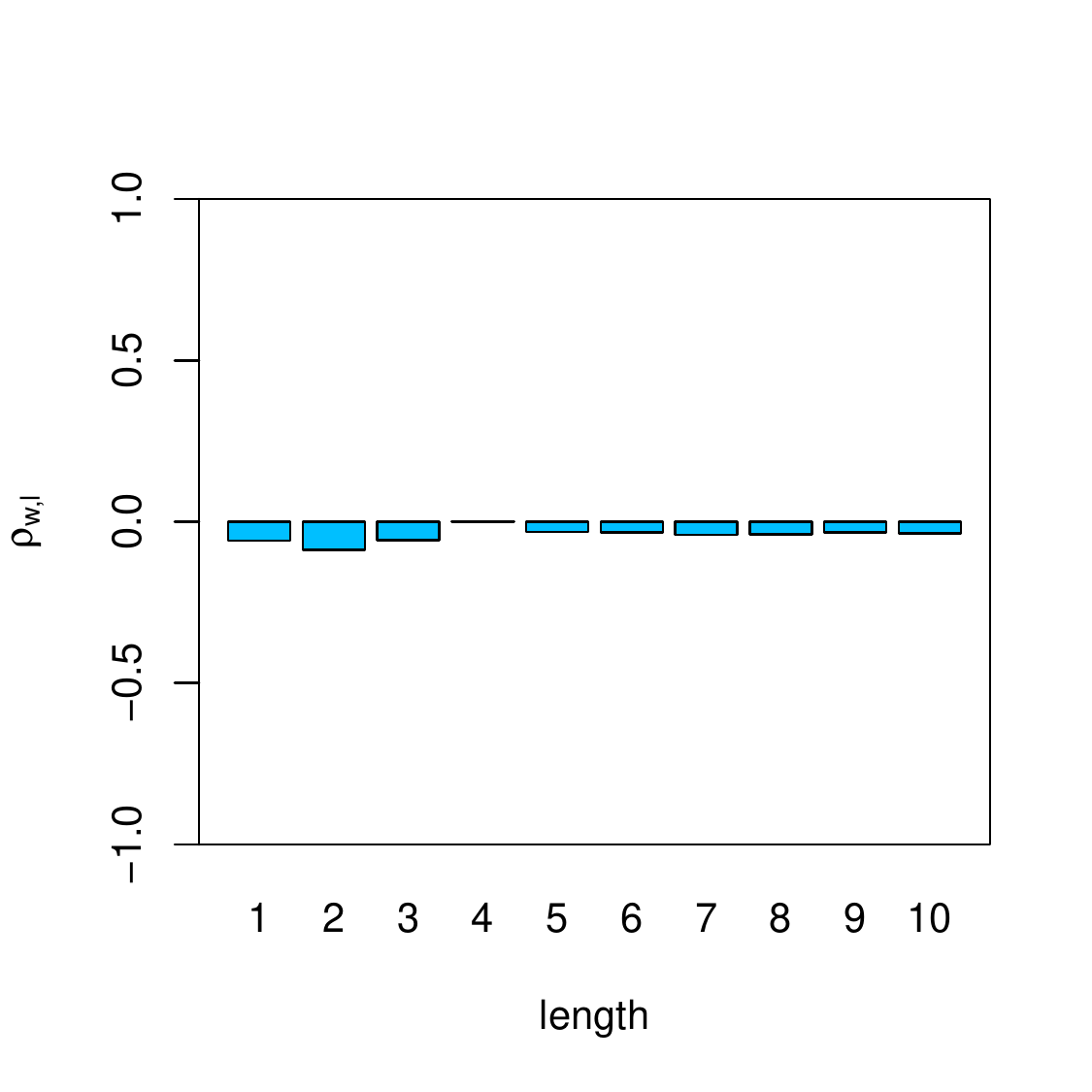}}\label{fig:G1UP}
\hspace{0.1\textwidth}
\subfigure[Assortativity through paths of $G_2$]{\includegraphics[width=0.4\textwidth]{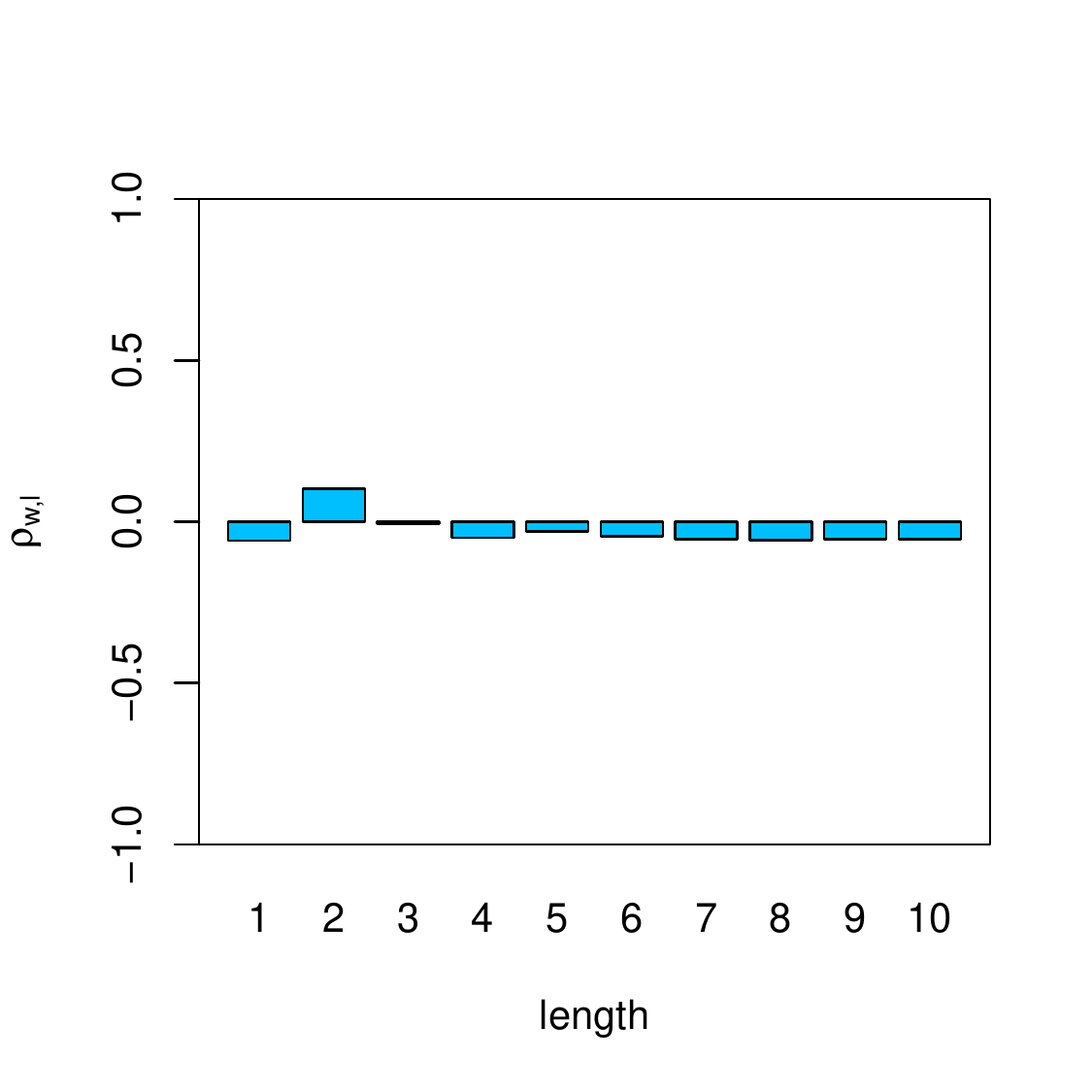}}\label{fig:G2UP}
\caption{Plot of higher order of assortativity measures of graphs $G_{1}$ and
$G_{2}$ depending on length $l$.}%
\label{fig:firstComparison}%
\end{figure}

Furthermore, the diagrams in Figures (\ref{fig:firstComparison}) from (a) to
(f) depict the plot of proposed assortativity measures, also shown in Table
\ref{tab:firstComparison}, for different values of lengths $l,$ allowing us to
focus on each measure separately. First of all, as we previously proved, all
measures for $l=1$ correspond to the Newman's index. Moreover, the
assortativity through random walk vanishes as $l$ approaches to infinity.

Looking at the values in Table \ref{tab:firstComparison}, referring to the
Newman's index, graphs are equally, slightly disassortative so, for both, high
degree nodes tend to be adjacent to low degree nodes. However, looking at the
assortativity beyond the nearest neighbors, the two graphs are different and
our measures are able to better capture the topological features related to
the assortativity.

Taking as an example $l=2,$ $G_{2}$ is certainly assortative, referring to all
the measures, as $\rho_{w,2}(G_{2})=0.3501,\rho_{p,2}(G_{2})=0.0951,$
$\rho_{up,2}(G_{2})=0.1024$, and this is due to the prevalence of existing
connections between similar degree nodes through 2 steps. On the contrary,
this effect is not present for $G_{1},$ that shows assortativity through
random walks of length 2 but not through 2-paths, as $\rho_{w,2}%
(G_{1})=0.1826,\rho_{p,2}(G_{1})=-0.0787,$ $\rho_{up,2}(G_{1})=-0.0864.$ This
result is consistent with the transitivity values, being $C(G_{2})$ higher
than $C(G_{1})$.

For $l=3,$ $G_{1}$ certainly becomes disassortative, referring to all the
measures, as $\rho_{w,3}(G_{1}),\rho_{p,3}(G_{1}),$ $\rho_{up,3}(G_{1})$ are
negative, prevailing the connections between high degree nodes with low degree
nodes through 3 steps. $G_{2}$ shows assortativity through random walks of
length $3$ and disassortativity through $3$-paths.

Looking at the diagrams in Figures (\ref{fig:firstComparison} a-b) , also for
$l=2,$ both graphs are assortative, in particular $G_{2}$ is more assortative
than $G_{1}$, whereas for $l=3,$ $G_{1}$ becomes disassortative and $G_{2}$
still assortative. In general, we can deduce that, for $G_{2},$ similar degree
nodes tend to be connected to each other by walks of any length $l>1,$ indeed
$\rho_{w,l}(G_{2})$ assumes positive signs. On the contrary, $\rho
_{w,l}\left(  G_{1}\right)  $ presents an alternating sequence of signs. It is
worth noting that, degree-based paths (Figures (\ref{fig:firstComparison}
c-f) do not significantly modify correlations and, for a given graph and a
given length, $\rho_{p,l}$ and $\rho_{up,l}$ are quite similar.

A similar analysis can be done for graphs $G_{3}$ and $G_{4}$ (see Figure
\ref{fig:secondComparisonGraphs}).

They share the same Newman's index $\rho(G_{3})=\rho
(G_{4})=0.0966$ so in this case they are both assortative. Classical network
indicators are $D(G_{3})=5,$ whereas $D(G_{4})=7;$ $L(G_{3})=2.5931$ whereas
$L(G_{4})=2.7448$. $C(G_{3})=0.1091$ whereas $C(G_{4})=0.2045$.

\begin{table}[ptb]
\caption{Higher order assortativity measures of graphs $G_{3}$ and $G_{4}.$}%
\label{tab:secondComparison}%
\centering
\begin{tabular}
[c]{r|rrr|rrr}%
$l$ & $\rho_{w,l}(G_{3})$ & $\rho_{p,l}(G_{3})$ & $\rho_{up,l} (G_{3})$ &
$\rho_{w,l}(G_{4})$ & $\rho_{p,l}(G_{4})$ & $\rho_{up,l}(G_{4})$\\\hline
1 & 0.0966 & 0.0966 & 0.0966 & 0.0966 & 0.0966 & 0.0966\\
2 & 0.2389 & 0.0198 & 0.0165 & 0.2307 & -0.1229 & -0.1037\\
3 & -0.0071 & -0.1209 & -0.1171 & 0.1019 & -0.0601 & -0.0449\\
4 & 0.0808 & -0.1192 & -0.1107 & 0.1258 & -0.0339 & -0.0374\\
5 & -0.0109 & -0.0656 & -0.0531 & 0.0816 & -0.0291 & -0.0298\\
6 & 0.0334 & -0.0565 & -0.0511 & 0.0850 & -0.0361 & -0.0374\\
7 & -0.0059 & -0.0445 & -0.0408 & 0.0623 & -0.0375 & -0.0403\\
8 & 0.0156 & -0.0438 & -0.0430 & 0.0615 & -0.0352 & -0.0378\\
9 & -0.0025 & -0.0465 & -0.0469 & 0.0472 & -0.0383 & -0.0394\\
10 & 0.0079 & -0.0491 & -0.0459 & 0.0457 & -0.0422 & -0.0420
\end{tabular}
\end{table}

Observing the values referred to graph $G_{3}$ in Table
(\ref{tab:secondComparison}), assortativity is also confirmed for $l=2,$
as $\rho_{w,2}(G_{3}),\rho_{p,2}(G_{3}),$ $\rho_{up,2}(G_{3})$ are positive,
whereas $G_{3}$ becomes disassortative for $l=3$.

On the contrary, assortativity is not confirmed for graph $G_{4}$ showing
assortativity through random walks of length $l>1$ but not through $l$-paths.
Notice that, in this case, results are not consistent with the transitivity
values, being now $C(G_{3})$ lower than $C(G_{4})$.

\begin{figure}[ptb]
\centering
\subfigure[Assortativity through random walks of $G_3$]{\includegraphics[width=0.4\textwidth]{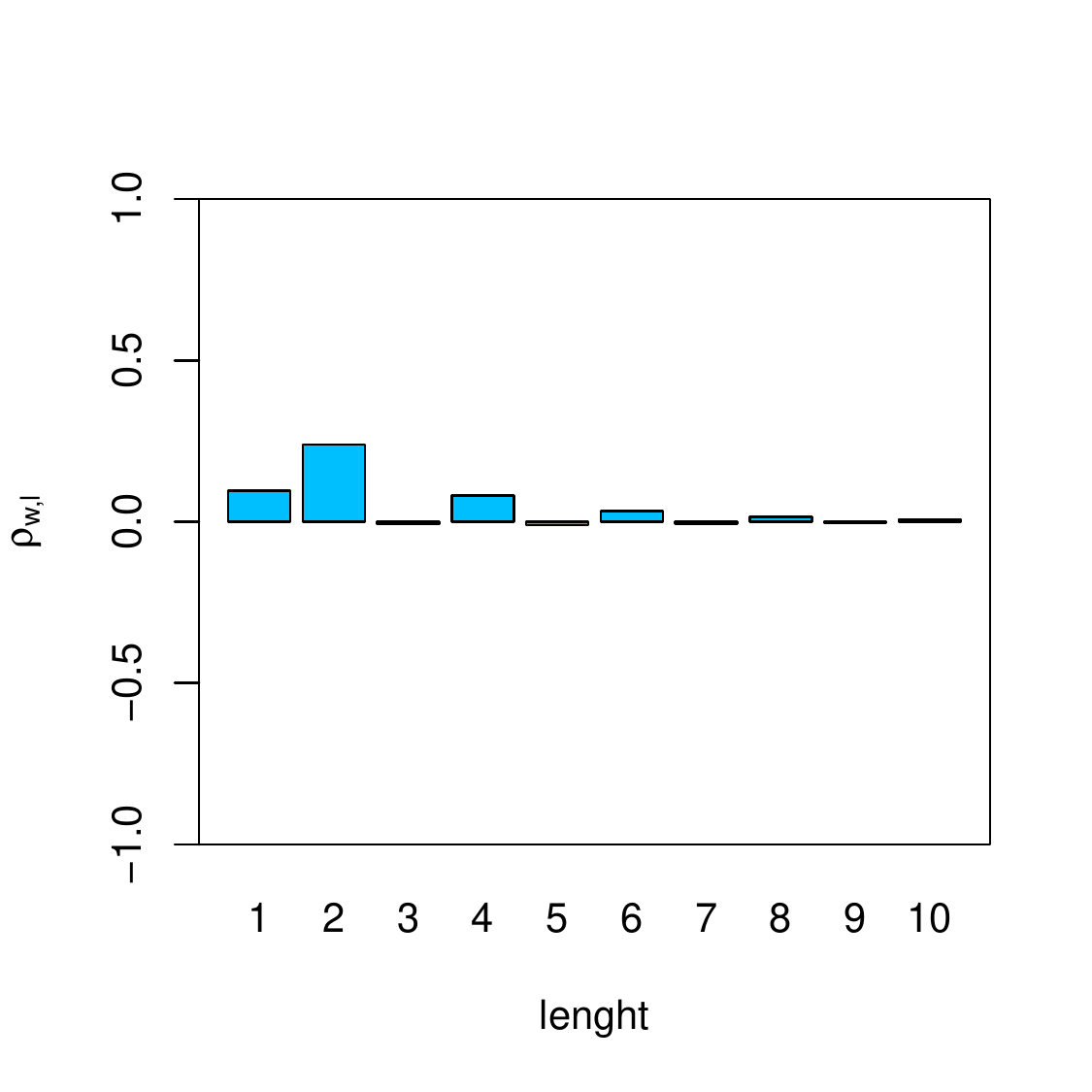}}\label{fig:G3RW}
\hspace{0.1\textwidth}
\subfigure[Assortativity through random walks of $G_4$]{\includegraphics[width=0.4\textwidth]{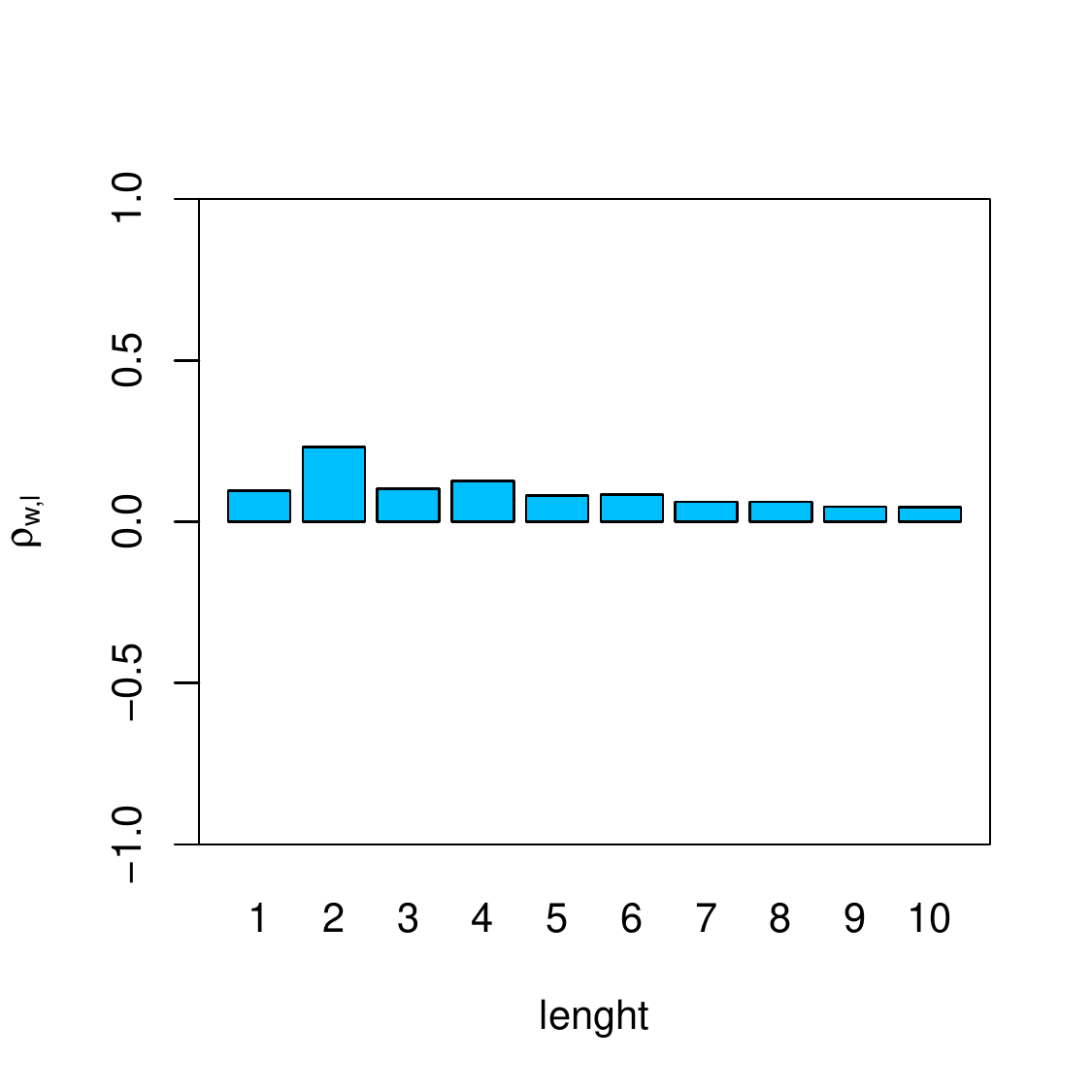}}\label{fig:G4RW}
\subfigure[Assortativity through degree-based paths of $G_3$]{\includegraphics[width=0.4\textwidth]{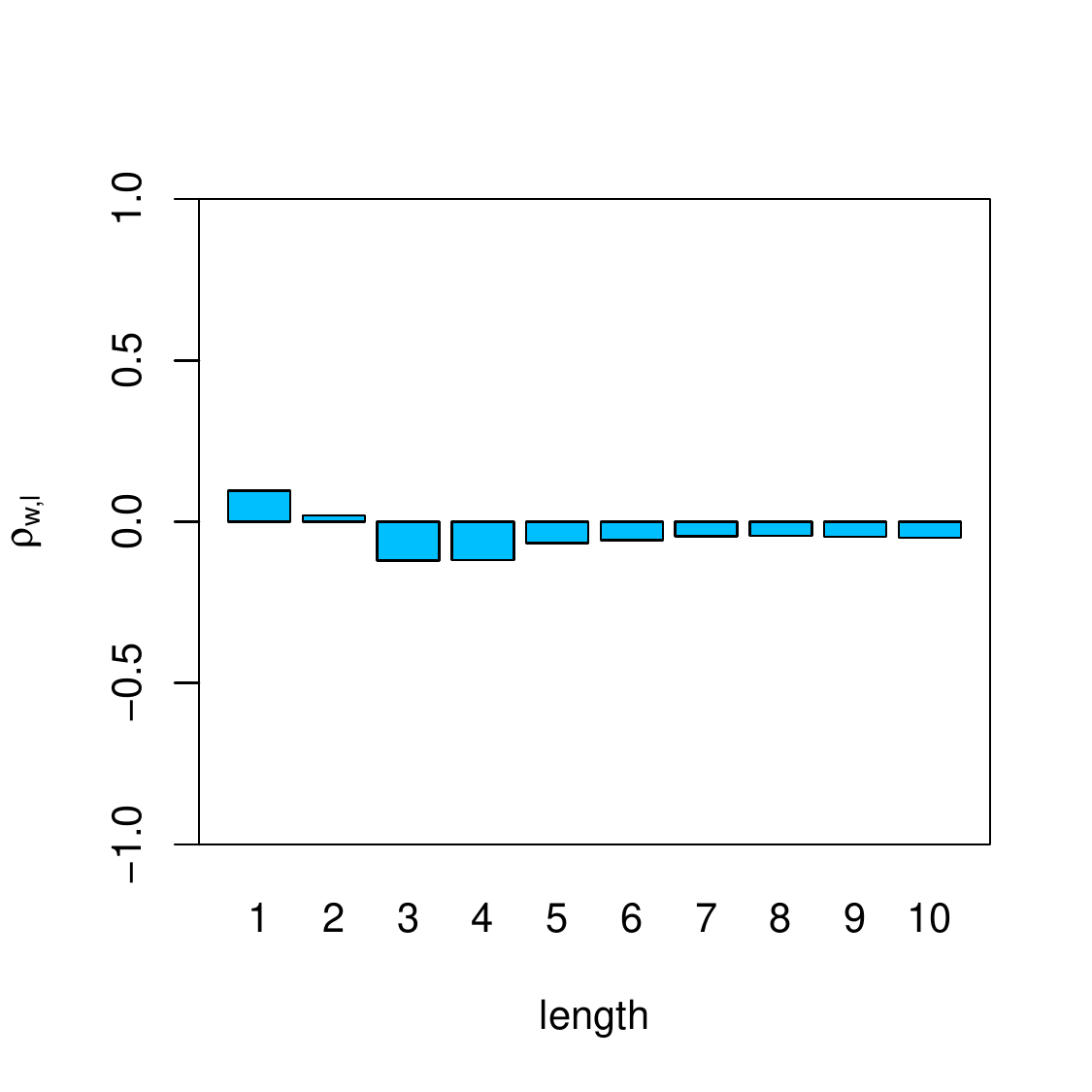}}\label{fig:G3WP}
\hspace{0.1\textwidth}
\subfigure[Assortativity through degree-based paths of $G_4$]{\includegraphics[width=0.4\textwidth]{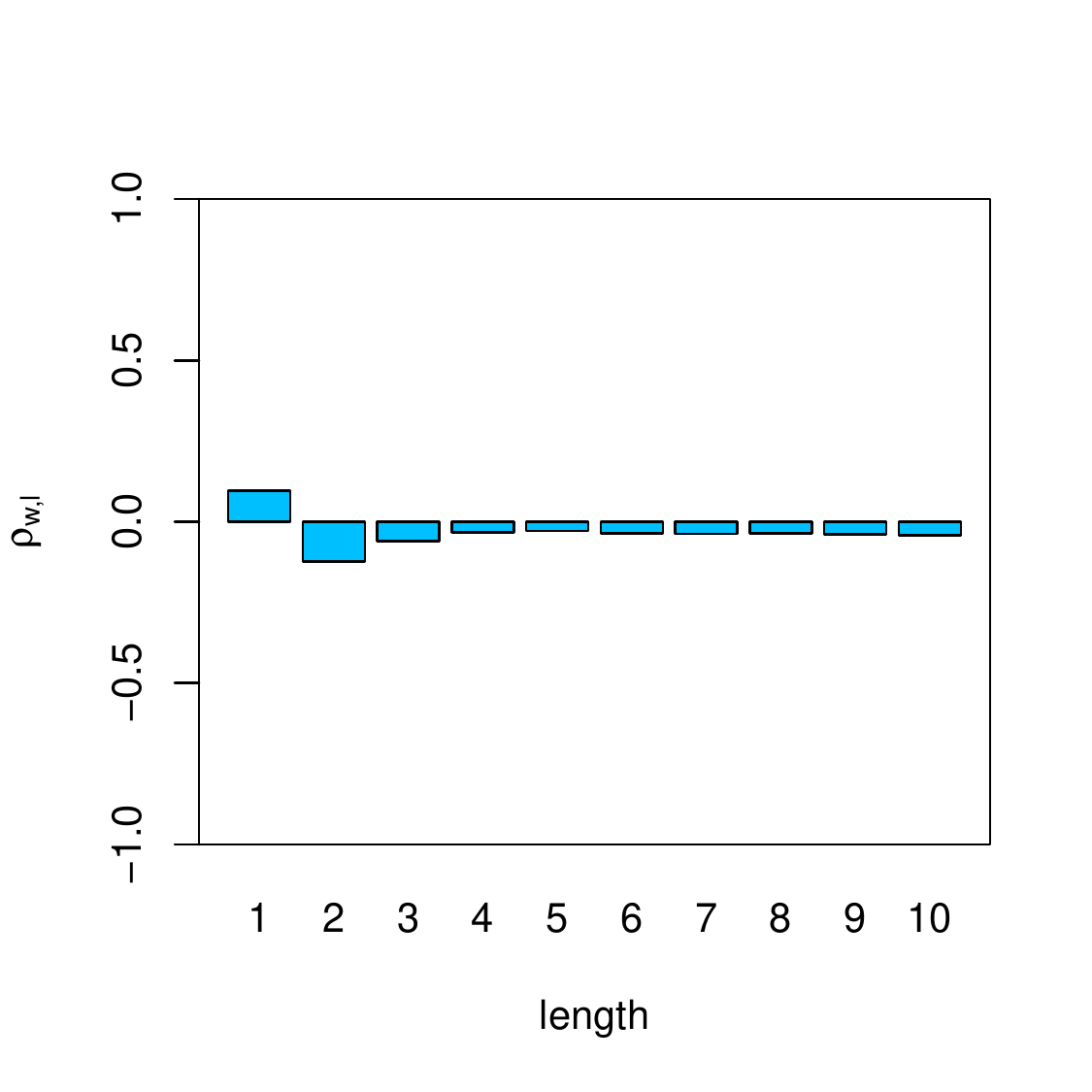}}\label{fig:G4WP}
\subfigure[Assortativity through paths of $G_3$]{\includegraphics[width=0.4\textwidth]{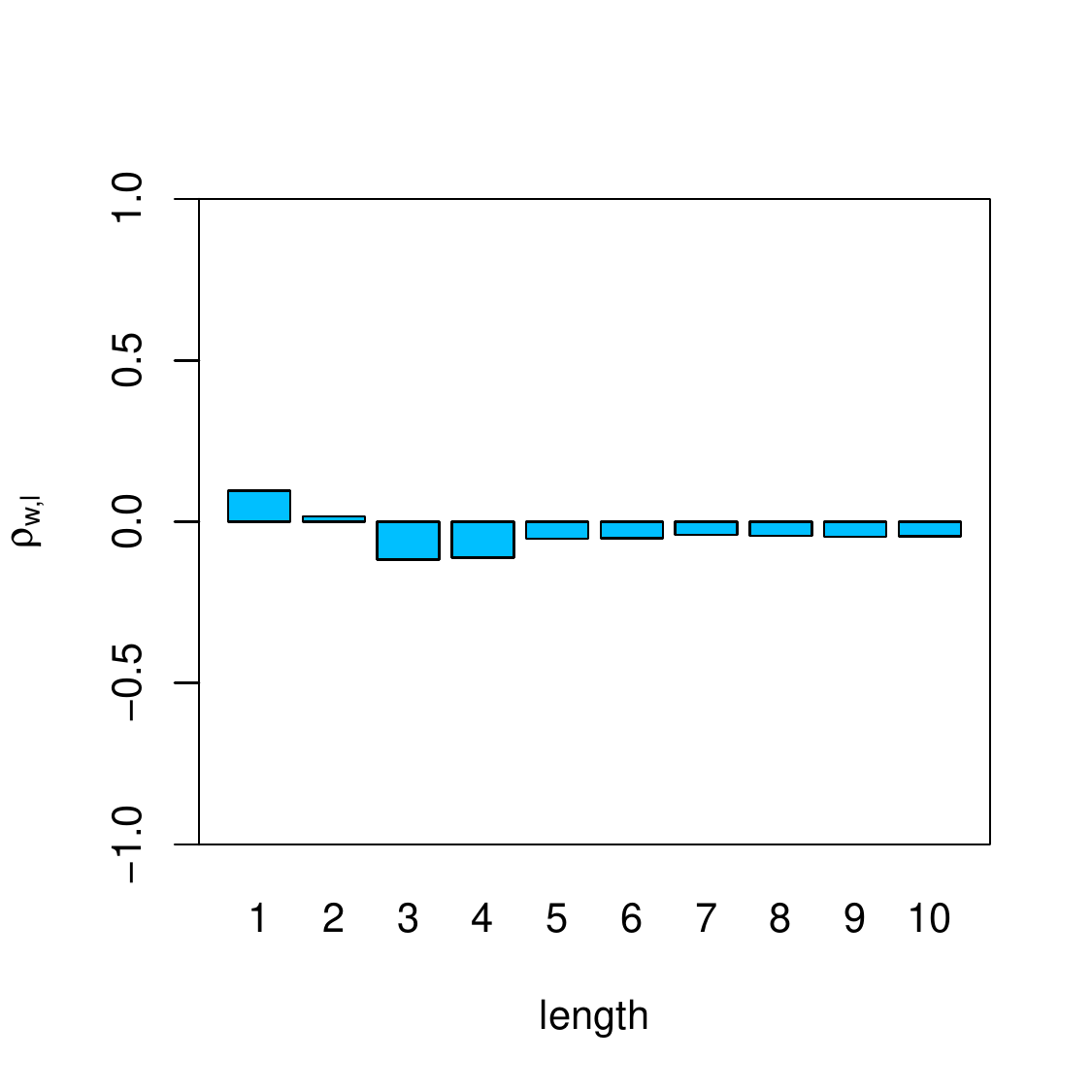}}\label{fig:G3UP}
\hspace{0.1\textwidth}
\subfigure[Assortativity through paths of $G_4$]{\includegraphics[width=0.4\textwidth]{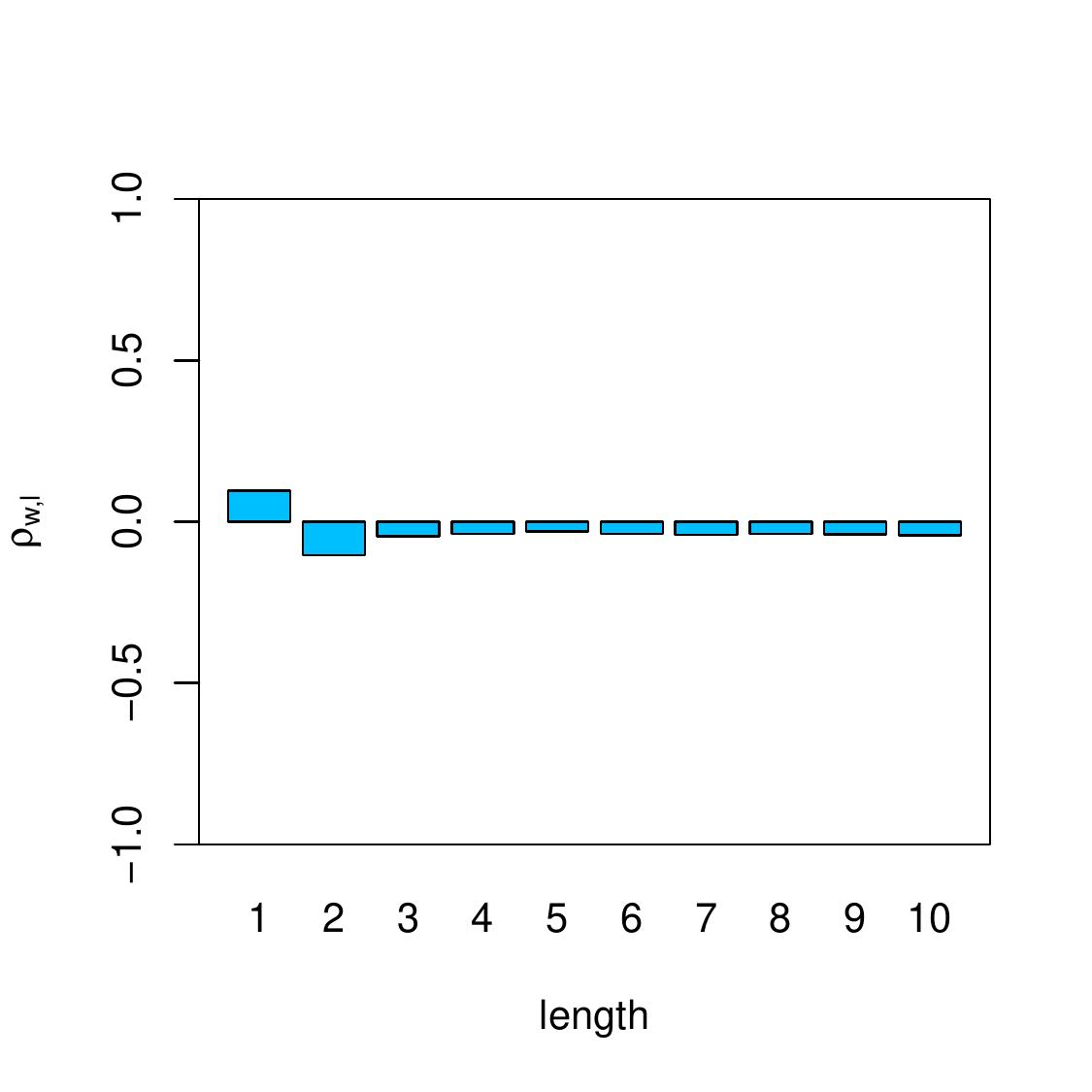}}\label{fig:G4UP}%
\caption{Assortativity measures of graphs $G_{3}$ and $G_{4}$}%
\label{fig:secondComparison}%
\end{figure}

Looking at Figure (\ref{fig:secondComparison} a-b), the measures of
assortativity through random walks of length 2 are similar, being $\rho
_{w,2}(G_{3})=0.2389$ and $\rho_{w,2}(G_{4})=0.2307$. Therefore, in order to
identify a graph through these measures it is convenient to consider them for all
lengths, taking into account the class of the measures as a whole. The behavior
becomes completely different taking into account longer random walks
($l>2$).\bigskip

Unlike the other indices we proposed, $\rho_{c,\alpha}(\cdot)$ (assortativity
through shortest paths) is not a function of path lengths therefore plots are
not provided. It is a measure that summarizes the graph assortativity
for different orders of lengths. In particular, distance between vertices is
used as weight to reduce the influence of distant vertices in the evaluation
of assortativity through all the couples of connected vertices. In this
example we evaluated the measure for $\alpha=1$. A larger $\alpha$ reduces the
relevance of vertices that are not adjacent, so that the Newman index is the
limit of $\rho_{c,\alpha}(\cdot)$ as $\alpha$ approaches infinity.
Therefore a comparison between $\rho_{c,1}(\cdot)$ and $\rho(\cdot)$ is useful
to understand the influence on the assortativity of vertices that are
connected but not adjacent.

The Newman index is equal and negative for both graphs $G_{1}$ and $G_{2}$
($\rho(G_{1})=\rho(G_{2})=-0.0584$) and the measure through shortest paths is
$\rho_{c,1}(G_{1})=-0.0327$ and $\rho_{c,1}(G_{2})=-0.0089$. Therefore we can
assert that both graphs are disassortative (vertices with large degrees tend
to be connected to vertices with low degrees) and that farther vertices reduce
the intensity of such relations.

The case of graphs $G_{3}$ and $G_{4}$ is more particular. Newman index
asserts that both graphs are equally assortative, $\rho(G_{3})=\rho
(G_{4})=0.0966$, but if we consider the synthetic measure through shortest
paths we observe that this relation vanishes and they become almost
disassortative, $\rho_{c,1}(G_{3})=-0.0074$ $\rho_{c,1}(G_{4})=-0.0035$. This
means that, even if vertices tend to be adjacent with other similar vertices,
when we also consider vertices connected by a larger distance there is no
linear relation between vertices degree because the measures are close to zero.

\section{Conclusions}

Using a unified approach, in this paper we have introduced high order assortativity based on paths, shortest paths and random walks. The analysis
has been performed for undirected and unweighted networks. Through simulations, we have shown that higher order assortativity can help to better reveal the network topology. The analysis can be possibly extended to weighted and directed networks.

\section*{Appendix 1}
\subsection*{Newman's assortativity index}

Here we report all mathematical details needed to obtain the Newman's formula
(\ref{Newman}) from expression (\ref{eq:rho_A}):
\begin{align*}
\rho &  =\frac{\mathbf{d}^{T}\left(  \frac{\mathbf{A}}{2M}-\frac
{\mathbf{d}\,\mathbf{d}^{T}}{4M^{2}}\right)  \,\mathbf{d}}{\mathbf{d}%
^{T}\,(\frac{\mathbf{D}}{2M}-\frac{\mathbf{d}\,\mathbf{d}^{T}}{4M^{2}%
})\,\mathbf{d}}=\frac{\frac{1}{2M}\mathbf{d}^{T}\mathbf{Ad}-\frac{1}{4M^{2}%
}\mathbf{d}^{T}\left(  \mathbf{d}\,\mathbf{d}^{T}\right)  \mathbf{d}%
}{\mathbf{d}^{T}\,\frac{\mathbf{D}}{2M}\mathbf{d}-\frac{1}{4M^{2}}%
(\mathbf{d}^{T}\mathbf{d}\,\mathbf{d}^{T}\mathbf{d)}\,}=\\
&  =\frac{\frac{1}{2M}\mathbf{d}^{T}\mathbf{Ad}-\frac{1}{4M^{2}}\left(
\mathbf{d}^{T}\mathbf{d}\right)  ^{2}}{\frac{1}{2M}\mathbf{d}^{T}%
\,\mathbf{D}\,\mathbf{d-}\frac{1}{4M^{2}}\left(  \mathbf{d}^{T}\mathbf{d}%
\right)  ^{2}}.
\end{align*}
\bigskip

Observe now that:%
\[
\mathbf{d}^{T}\mathbf{Ad}=\sum_{i}d_{i}\left(  \sum_{j}a_{ij}d_{j}\right)
=\sum_{i}\sum_{j}d_{i}d_{j}a_{ij}=2\sum_{i\sim j}d_{i}d_{j}
\]

since in the left-hand side, the sum is over all possible couples $\left(
i,j\right)  ,$ whereas on the right-hand side, the sum is over all adjacent
couples $\left(  i,j\right)  $ and nodes $i$ and $j$ are counted twice.

Now it is easy to check the following chain of equalities:%
\[
\mathbf{d}^{T}\mathbf{d=}\sum_{i}d_{i}^{2}=\sum_{i\sim j}\left(  d_{i}%
+d_{j}\right)  =\sum_{i}\sum_{j}\frac{1}{2}\left(  d_{i}+d_{j}\right)
a_{ij}.
\]

Indeed, the summation of $\left(  d_{i}+d_{j}\right)  $ is over all couples of
adjacent nodes, so that every term $d_{i}$ appears as much times as its
degree, i.e. $d_{i}$ times.

Similar argument leads to the following chain of equality:
\[
\mathbf{d}^{T}\,\mathbf{D}\,\mathbf{d=}\sum_{i}d_{i}^{3}=\sum_{i\sim j}\left(
d_{i}^{2}+d_{j}^{2}\right)  =\sum_{i}\sum_{j}\frac{1}{2}\left(  d_{i}%
^{2}+d_{j}^{2}\right)  a_{ij.}
\]

Then the numerator becomes:%
\begin{align*}
\frac{1}{2M}\mathbf{d}^{T}\mathbf{Ad}-\frac{1}{4M^{2}}\left(  \mathbf{d}%
^{T}\mathbf{d}\right)  ^{2}  &  =\frac{2\sum_{i\sim j}d_{i}d_{j}}{2M}%
-\frac{\left(  \sum_{i\sim j}\left(  d_{i}+d_{j}\right)  \right)  }{4M^{2}%
}^{2}=\\
&  =\frac{1}{M}\sum_{i\sim j}d_{i}d_{j}-\left[  \frac{1}{2M}\sum_{i\sim
j}\left(  d_{i}+d_{j}\right)  \right]  ^{2}=\\
&  =\frac{1}{2M}\sum_{i}\sum_{j}d_{i}d_{j}a_{ij}-\left[  \frac{1}{4M}\sum
_{i}\sum_{j}\left(  d_{i}+d_{j}\right)  a_{ij}\right]  ^{2}%
\end{align*}

The denominator is:%
\begin{align*}
\frac{1}{2M}\mathbf{d}^{T}\,\mathbf{D}\,\mathbf{d-}\frac{1}{4M^{2}}\left(
\mathbf{d}^{T}\mathbf{d}\right)  ^{2}  &  =\frac{1}{2M}\sum_{i\sim j}\left(
d_{i}^{2}+d_{j}^{2}\right)  -\frac{\left(  \sum_{i\sim j}\left(  d_{i}%
+d_{j}\right)  \right)  }{4M^{2}}^{2}=\\
&  =\frac{1}{2M}\sum_{i\sim j}\left(  d_{i}^{2}+d_{j}^{2}\right)  -\left[
\frac{1}{2M}\sum_{i\sim j}\left(  d_{i}+d_{j}\right)  \right]  ^{2}=\\
&  =\frac{1}{4M}\sum_{i}\sum_{j}\left(  d_{i}^{2}+d_{j}^{2}\right)
a_{ij}-\left[  \frac{1}{4M}\sum_{i}\sum_{j}\left(  d_{i}+d_{j}\right)
a_{ij}\right]  ^{2}%
\end{align*}

yielding to the final formula:\bigskip%

\[
\rho=\frac{\frac{1}{2M}\sum_{i}\sum_{j}d_{i}d_{j}a_{ij}-\left[  \frac{1}%
{4M}\sum_{i}\sum_{j}\left(  d_{i}+d_{j}\right)  a_{ij}\right]  ^{2}}{\frac
{1}{4M}\sum_{i}\sum_{j}\left(  d_{i}^{2}+d_{j}^{2}\right)  a_{ij}-\left[
\frac{1}{4M}\sum_{i}\sum_{j}\left(  d_{i}+d_{j}\right)  a_{ij}\right]  ^{2}}.
\]

and its equivalent forms:%
\begin{align*}
\rho &  =\frac{\frac{1}{M}\sum_{i\sim j}d_{i}d_{j}-\left[  \frac{1}{2M}%
\sum_{i\sim j}\left(  d_{i}+d_{j}\right)  \right]  ^{2}}{\frac{1}{2M}%
\sum_{i\sim j}\left(  d_{i}^{2}+d_{j}^{2}\right)  -\left[  \frac{1}{2M}%
\sum_{i\sim j}\left(  d_{i}+d_{j}\right)  \right]  ^{2}}=\\
&  =\frac{\sum_{i\sim j}d_{i}d_{j}-\left[  \sum_{i}\frac{1}{2}d_{i}%
^{2}\right]  ^{2}/M}{\sum_{i}\frac{1}{2}d_{i}^{3}-\left[  \sum_{i}\frac{1}%
{2}d_{i}^{2}\right]  ^{2}/M}.
\end{align*}

\section*{Appendix 2}
\subsection*{Proof of Theorem 1}

\emph{ Let $G=(V,E)$ be a graph with adjacency matrix $\mathbf{A}$ and degree
sequence $\mathbf{d.}$ Let $\mathbf{P}$ be the transition matrix of a Markov
chain on $G=(V,E)$. If $\mathbf{P}$ is primitive, the assortativity of order
$l,$ $\rho_{w,l}$ vanishes as \thinspace$l\rightarrow\infty.$}

\begin{proof}
Let us now consider a Markov chain on $G=(V,E)$ with adjacency matrix
$\mathbf{A}$ and degree sequence $\mathbf{d}$ (see \cite{Grinstead Snell
1997}) and let $\mathbf{P}$ be the transition matrix. First of all,
$\mathbf{P}^{l}$ provides the partial probability distributions of being at
the $j$-th state of the Markov chain after $l$ steps starting from the $i$-th
state, then $\mathbf{E}_{w,l}=\mathbf{D}_{\mathbf{q}}\mathbf{P}^{l},$ i.e.
the  probabilities that a walk randomly chosen from $E_{w,l}$ connects
vertices $i$  and $j$ can be obtained by multiplying each partial distribution
by the  probability to be in the $i$-th state. Observe that also for $l>1$ it
holds  $\mathbf{q}=\frac{1}{2m}\mathbf{d,}$ for a well known property of
Markov chain  on undirected graphs.

As a consequence, the measure of assortativity of order $l$ is:
\begin{equation}
\rho_{w,l}=\frac{\mathbf{d}^{T}\left(  \mathbf{D}_{\mathbf{q}}\mathbf{P}
^{l}-\mathbf{q}\mathbf{q}^{T}\right)  \mathbf{d}}{\mathbf{d}^{T}\left(
\mathbf{D}_{\mathbf{q}}-\mathbf{q}\mathbf{q}^{T}\right)  \mathbf{d}}
\label{eq:rhol}%
\end{equation}

Recalling that $\mathbf{P=D}^{-1}\mathbf{A}$ and $\mathbf{D}_{\mathbf{q}
}=\frac{1}{2m}\mathbf{D,}$ (\ref{eq:rhol}) can be rewritten as:
\[
\rho_{w,l}=\frac{\mathbf{d}^{T}\left(  \frac{\mathbf{D}\left(  \mathbf{D}
^{-1}\mathbf{A}\right)  ^{l}}{2m}-\frac{\mathbf{dd}^{T}}{4m^{2}}\right)
\mathbf{d}}{\mathbf{d}^{T}\left(  \frac{\mathbf{D}}{2m}-\frac{\mathbf{dd}^{T}
}{4m^{2}}\right)  \mathbf{d}}.
\]

Given the connectedness of graph $G,$ observe that the vector $\mathbf{q}$ is
the unique stationary distribution. Being the matrix $\mathbf{P}$ is
primitive, then $\lim_{l\rightarrow\infty}\mathbf{P}^{l}$ \ exists and:
\[
\lim_{l\rightarrow+\infty}\mathbf{P}^{l}=\left[
\begin{array}
[c]{cccc}%
\mathbf{q} & \mathbf{q} & \ldots & \mathbf{q}%
\end{array}
\right]  ^{T}=\left[
\begin{array}
[c]{cccc}%
\frac{\mathbf{d}}{2m} & \frac{\mathbf{d}}{2m} & \ldots & \frac{\mathbf{d}}{2m}%
\end{array}
\right]  ^{T}=\frac{1}{2m}\mathbf{1d}\,^{T};
\]
we will call this matrix $\mathbf{P}^{\infty}.$ Observing that
\begin{equation}
\mathbf{D}_{\mathbf{q}}\mathbf{P}^{\infty}=\frac{\mathbf{d}\mathbf{d}^{T}
}{4m^{2}}=\mathbf{q}\mathbf{q}^{T} \label{E_inf}%
\end{equation}
equation (\ref{eq:rhol}) can be rewritten as:
\begin{equation}
\rho_{w,l}=\frac{\mathbf{d}^{T}\left[  \mathbf{D}_{\mathbf{q}}\left(
\mathbf{P}^{l}-\mathbf{P}^{\infty}\right)  \right]  \mathbf{d}}{\mathbf{d}
^{T}\left[  \mathbf{D}_{\mathbf{q}}\left(  \mathbf{I}-\mathbf{P}^{\infty
}\right)  \right]  \mathbf{d}}%
\end{equation}
then the assortativity of order $l$ vanishes as \thinspace$l\rightarrow
\infty.$
\end{proof}

\subsection*{Proof of Theorem 2}

\emph{ Let $G=(V,E)$ be a simple connected graph. The coefficient
$\rho_{c,\alpha}$ tends to $\rho$ as $\alpha$ tends infinity.}

\begin{proof}
The matrix $\mathbf{H}_{\alpha}$ contains the reciprocal of $\alpha$-power of
the distances between all couples of distinct nodes, adjacents or not. Being
the graph $G$ connected, at least one path between every couple of nodes $i$
and $j,$ $(i\not =j)$ exists, then $d(i,j)\geq1.$ In particular, $d\left(
i,j\right)  =1$ only if $i$ and $j$ are adjacents and in this case
$dist(i,j)^{-\alpha}=1$ for all $\alpha,$ otherwise, $\lim_{\alpha
\rightarrow+\infty}dist(i,j)^{-\alpha}=0$. Then, when $\alpha$ approaches to
infinity, $\mathbf{H}_{\alpha}\rightarrow\mathbf{A,}$ $h=\sum_{i}\sum
_{j}h_{ij}\rightarrow2m,$ $\mathbf{q}\rightarrow\frac{\mathbf{d}}{2m}$ and
finally $\rho_{c,\alpha}\rightarrow\rho.$
\end{proof}

\end{document}